\documentclass[journal]{IEEEtran}
\usepackage{stfloats}

\usepackage{enumerate}
\usepackage{algorithm,algpseudocode}
\usepackage{setspace,amsmath,latexsym,cite,amssymb,epsfig,amsfonts}
\usepackage{url,cite}
\usepackage{graphicx}
\usepackage{psfrag}
\usepackage{footmisc}
\usepackage{multirow}
\usepackage{color}
\usepackage{multicol}
\usepackage{mathtools}
\usepackage{subcaption}
\usepackage{graphicx}
\usepackage{amssymb}
\usepackage{amsmath}
\usepackage{epstopdf}
\usepackage{geometry}
\usepackage{float}
\usepackage{balance}
\usepackage{makecell}
\usepackage{changepage}

\algnewcommand\algorithmicforeach{\textbf{for}}
\algdef{S}[FOR]{For}[1]{\algorithmicforeach\ #1\ \algorithmicdo}

\twocolumn

\makeatother

        \makeatletter
        \def\fps@eqnfloat{!t}
        \def\ftype@eqnfloat{4}
        
        \newenvironment{eqnfloat*}
               {\@dblfloat{eqnfloat}}
               {\end@dblfloat}
        \makeatother
\allowdisplaybreaks[4]

\title{\huge Fair Resource Allocation For Hierarchical Federated Edge Learning in Space-Air-Ground Integrated Networks via\\Deep Reinforcement Learning with Hybrid Control}
\author{Chong Huang, \IEEEmembership{Member, IEEE}, Gaojie Chen, \IEEEmembership{Senior Member, IEEE}, Pei Xiao, \IEEEmembership{Senior Member, IEEE},\\Jonathon A. Chambers, \IEEEmembership{Fellow, IEEE}, and Wei Huang
\thanks{C. Huang, G. Chen and P. Xiao are with 5GIC \& 6GIC, Institute for Communication Systems (ICS), Home for 5GIC \& 6GIC, University of Surrey, Guildford, GU2 7XH, United Kingdom. Email: \{chong.huang, gaojie.chen, p.xiao\}@surrey.ac.uk. (Corresponding author: G. Chen)}
\thanks{Jonathon A. Chambers is with the School of Engineering, University of Leicester, Leicester, LE1 7RU, United Kingdom. Email: jonathon.chambers@leicester.ac.uk.}
\thanks{W. Huang is with School of Flexible Electronics (SoFE) \& State Key Laboratory of Optoelectronic Materials and Technologies, Sun Yat-sen University, Guangdong, 510220, China. Email: huangw323@mail.sysu.edu.cn}
}

\begin{document}
\captionsetup[figure]{name={Fig.},labelsep=period}

\begin{singlespace}
\maketitle

\end{singlespace}

\thispagestyle{empty}
\begin{abstract}
The space-air-ground integrated network (SAGIN) has become a crucial research direction in future wireless communications due to its ubiquitous coverage, rapid and flexible deployment, and multi-layer cooperation capabilities. However, integrating hierarchical federated learning (HFL) with edge computing and SAGINs remains a complex open issue to be resolved. This paper proposes a novel framework for applying HFL in SAGINs, utilizing aerial platforms and low Earth orbit (LEO) satellites as edge servers and cloud servers, respectively, to provide multi-layer aggregation capabilities for HFL. The proposed system also considers the presence of inter-satellite links (ISLs), enabling satellites to exchange federated learning models with each other. Furthermore, we consider multiple different computational tasks that need to be completed within a limited satellite service time. To maximize the convergence performance of all tasks while ensuring fairness, we propose the use of the distributional soft-actor-critic (DSAC) algorithm to optimize resource allocation in the SAGIN and aggregation weights in HFL. Moreover, we address the efficiency issue of hybrid action spaces in deep reinforcement learning (DRL) through a decoupling and recoupling approach, and design a new dynamic adjusting reward function to ensure fairness among multiple tasks in federated learning. Simulation results demonstrate the superiority of our proposed algorithm, consistently outperforming baseline approaches and offering a promising solution for addressing highly complex optimization problems in SAGINs.
\end{abstract}

\begin{IEEEkeywords}
Space-air-ground integrated network, hierarchical federated Learning, federated edge learning, deep reinforcement learning, satellite communications, unmanned aerial vehicle.
\end{IEEEkeywords}

\section{Introduction}
\subsection{Background}
With the advancement of wireless communications and artificial intelligence (AI) technologies, there has been a rapid increase in the demand for intelligent wireless networks in recent years \cite{9916263}. Many intelligent applications in wireless networks are trained based on local data from wireless nodes, which are stored in a distributed manner across these nodes. In traditional centralized machine learning methods, it is necessary to transfer vast amounts of data to the cloud or data centers for processing, this approach not only consumes a significant amount of wireless bandwidth but also leads to considerable latency \cite{10236519}. Moreover, data privacy and information security issues have become a major concern in the current digital era, because many applications involve the processing of sensitive personal information \cite{9141214}. To address the above issues, federated learning has emerged as a very popular machine learning framework in wireless communications \cite{9530714}. Within the federated learning framework, local devices in wireless networks train their models using their own local data, and only share their model parameters rather than their training data to protect data privacy. Furthermore, federated learning significantly reduces the amount of data that needs to be transmitted over wireless communications, thus saving wireless communication resources.

On the other hand, satellite networks have attracted much attention in wireless communications due to their capability to provide global coverage and seamless connections \cite{9508471}. In recent years, satellite network projects like Starlink \cite{McDowell_2020}, OneWeb \cite{foreman2017large}, and O3b \cite{8473417} have been rapidly advanced to provide global communication services, especially offering connections to remote areas. Thus, satellite communication has become an important component of future wireless communications. Furthermore, considering the strong demand for communication latency and rates in sixth-generation (6G) communications, the integration of satellite networks with terrestrial communication networks has emerged as a very promising design paradigm \cite{9502642}. In particular, space-air-ground integrated networks (SAGINs) seamlessly integrate satellites, aerial platforms, and terrestrial communication nodes, where satellites provide ubiquitous communication connections globally, aerial platforms offer rapid and flexible edge access capabilities for terrestrial networks, and terrestrial networks provide localized communication connections as the communication backbone. This integrated Space-Air-Ground framework design offers unprecedented prospects for future communications.

Moreover, edge computing was emerged to meet the growing demand for low-latency and high-bandwidth services by leveraging the computational and storage capabilities of terminal devices and edge servers \cite{7901477}. In edge computing, computing tasks are transferred from terminals to edge servers to reduce service latency in the wireless network \cite{9950554}. However, traditional edge computing requires the exchange of local data, occupying a large amount of communication bandwidth and raising issues of personal data privacy \cite{9060868}. Hence, hierarchical federated learning (HFL) was developed to address these issues via combining federated learning and edge computing \cite{9377005}. In HFL, models from terminals are aggregated multiple times at edge servers before being aggregated in the cloud server to efficiently utilize edge network resource while protecting data privacy. Furthermore, the hierarchical training process in HFL naturally aligns with the multi-level communication architecture of SAGINs \cite{10113881}. The global coverage capability and flexible aerial deployment of SAGINs enable more ground users to participate in HFL to significantly increase the diversity and performance of task training.

\subsection{Related Work}\label{sec:RW}
Federated learning is emerging as a paradigm for providing artificial intelligence services in wireless communications, and numerous current studies have investigated its performance within wireless networks \cite{9210812,10000935,9895272,10209589}. In \cite{9210812}, the Hungarian algorithm was introduced to optimize the user selection and resource block allocation to improve the federated learning aggregation performance in wireless networks. A distributed resource allocation method was proposed in \cite{10000935} to minimize the energy consumption and latency of federated learning in transmission control protocol/internet protocol (TCP/IP) networks. In \cite{9895272}, the delay performance was analyzed to improve the federated learning performance and the energy efficiency in wireless vehicular networks. To reduce the convergence time of federated learning in wireless communications, the authors of \cite{10209589} analyzed the effect of quantization errors and limited wireless resources in federated learning and proposed a model-based reinforcement learning algorithm to adjust the local model quantization and the number of local users in federate learning.

On the other hand, SAGIN has become a pivotal research direction for future wireless communications \cite{9177315,9062531,9701875,10440193}. In \cite{9177315}, the outage performance of the relaying scenario was analyzed with a closed form for SAGIN. To strike the balance between computational resource consumption and communication resource consumption, a heuristic greedy algorithm was proposed to design the virtual network functions and data routing in SAGINs. In \cite{9701875}, a queuing game model based method was introduced to enhance the network throughput and reduce the service delay in SAGINs. To minimize the energy consumption and latency for cloud and edge computing in SAGINs, a decision-assisted reinforcement learning algorithm was proposed in \cite{10440193} to optimize the resource allocation. Considering a SAGIN's global coverage capability and rapid dynamic deployment, it enables a more flexible participation of ground users in federated learning tasks, thus federated learning within SAGIN has been extensively researched in recent studies \cite{9982621,10149205}. In \cite{9982621}, a novel topology-aware framework was designed to speed up the convergence for federated learning in SAGINs. To jointly consider the requirements of adaptivity, communication efficiency and model security in federated learning, a tensor-empowered federated learning framework was proposed to meet these three requirements in SAGINs \cite{10149205}. The existing research on federated learning within SAGINs has focused on two-layer federated learning architectures, failing to fully leverage the potential performance capabilities of nodes within the SAGIN framework.

However, the conventional two-layer federated learning framework design often fails to meet the demands in scenarios with dispersed user locations and unstable communication channels \cite{10059225}. Moreover, the integration of federated learning with network edges in recent years enables low-latency edge intelligence at the data generation source. Therefore, multi-layer HFL with network edges has emerged as a highly promising research direction in wireless communications \cite{10059225,10089406,10155155}. In \cite{10059225}, a semi-decentralized framework was proposed to utilize the edge servers to enhance the convergence performance of federated learning. To improve the convergence performance and reduce the round delay in federated learning, a hierarchical aggregation strategy was proposed to optimize the edge server scheduling and bandwidth allocation in wireless communications \cite{10089406}. In \cite{10155155}, a hybrid deep reinforcement learning (DRL) algorithm was utilized to optimize the resource allocation strategy for fast convergence of federated learning in wireless communications. However, two significant challenges remain in the existing works: Firstly, the deployment of HFL utilizing space, air, and ground nodes within SAGINs as edge and cloud servers remains an area that has not been investigated. Secondly, while existing federated learning research has explored personalized multi-task learning with non-independent and identically distributed (non-i.i.d.) data \cite{NEURIPS2021_82599a4e}, achieving balance in the convergence performance of multiple completely unrelated tasks which are repeatedly distributed in multiple users within constrained timeframes, such as the service times of SAGIN environments, continues to be an unaddressed issue.

\subsection{Motivation and Contributions}
In this paper, we explore the integration of HFL in SAGINs. To the best of the authors' knowledge, we address a research gap as the existing works have not considered the utilization of the HFL framework in SAGIN to leverage the performance of various nodes and balance the convergence performance of different tasks within a limited satellite service time. Our study also encompasses the investigation of the impact of dynamic deployment and user association within SAGIN on HFL. To optimize the communication rounds within HFL and balance the final convergence performance of different tasks, dynamic planning of unmanned aerial vehicles (UAVs), user association, access between satellites and UAVs, the numbers of local convergence iterations and the weights of models in aggregations are crucial for HFL in SAGINs. However, traditional algorithms struggle to adapt to dynamically changing communication environments. Thus, we propose a hybrid action space DRL-based algorithm to address this challenge. The main contributions of this paper are summarized as follows.

\begin{itemize}
  \item We first propose a novel framework of HFL in SAGINs, where UAVs are considered as edge servers, LEO satellites as cloud servers, and the use of ISL for aggregation between multiple satellites and transferring aggregated models. To counteract channel fading effects in ground-aerial communications, we introduce trajectory planning for UAVs. Moreover, we consider the time constraints of LEO satellite access to aerial platforms, which is crucial for completing tasks within a limited satellite service time in practical applications and has not been taken into account by most studies.

  \item Our objective is to accomplish multiple federated learning tasks within a finite satellite service time while ensuring fairness among different task performances. Thus, we formulate a complex optimization problem in our proposed system, which includes UAV trajectory planning, dynamic adjustment of ground user-UAV pairing (uplink and downlink), UAV-satellite pairing (uplink and downlink), final aggregation selection between satellites, and optimization of weights in edge and cloud aggregation.

  \item We introduce the distributional soft-actor-critic (DSAC) algorithm, leveraging its ability to consider long-term returns and mitigate the overestimation issue of Q-values using return distributions. Moreover, due to the large number of coupled discrete and continuous optimization variables in the proposed problem, we propose a decoupling and recoupling algorithm to enhance the DSAC's performance. Finally, to ensure fairness, we design a dynamic adjusting reward function for the proposed algorithm to mitigate training progress deviations among different tasks.

  \item Simulation results demonstrate the superior performance of our proposed algorithm, compared to several benchmarks. Through the design of a hybrid action space, the proposed DSAC algorithm fully learns the proposed system and obtains corresponding solutions. Moreover, the use of the dynamic adjusting reward design ensures fairness among multiple tasks.
\end{itemize}

The rest of this paper is summarized as follows: The system model including the communication model, satellite coverage model, federated learning framework and problem formulation are introduced in Section \ref{sec:sm}. In Section \ref{sec:DSAC}, a hybrid action space DRL algorithm with a dynamic adjusting reward function is proposed. Section \ref{sec:sim} verifies the performance of the proposed method for HFL in SAGINs via simulations. Finally, Section \ref{sec:con} summarizes this work.

\section{System Model and Problem Formulation} \label{sec:sm}
\subsection{System Model}
\begin{figure}[t!]
  \centering
  \centerline{\includegraphics[width=0.46\textwidth]{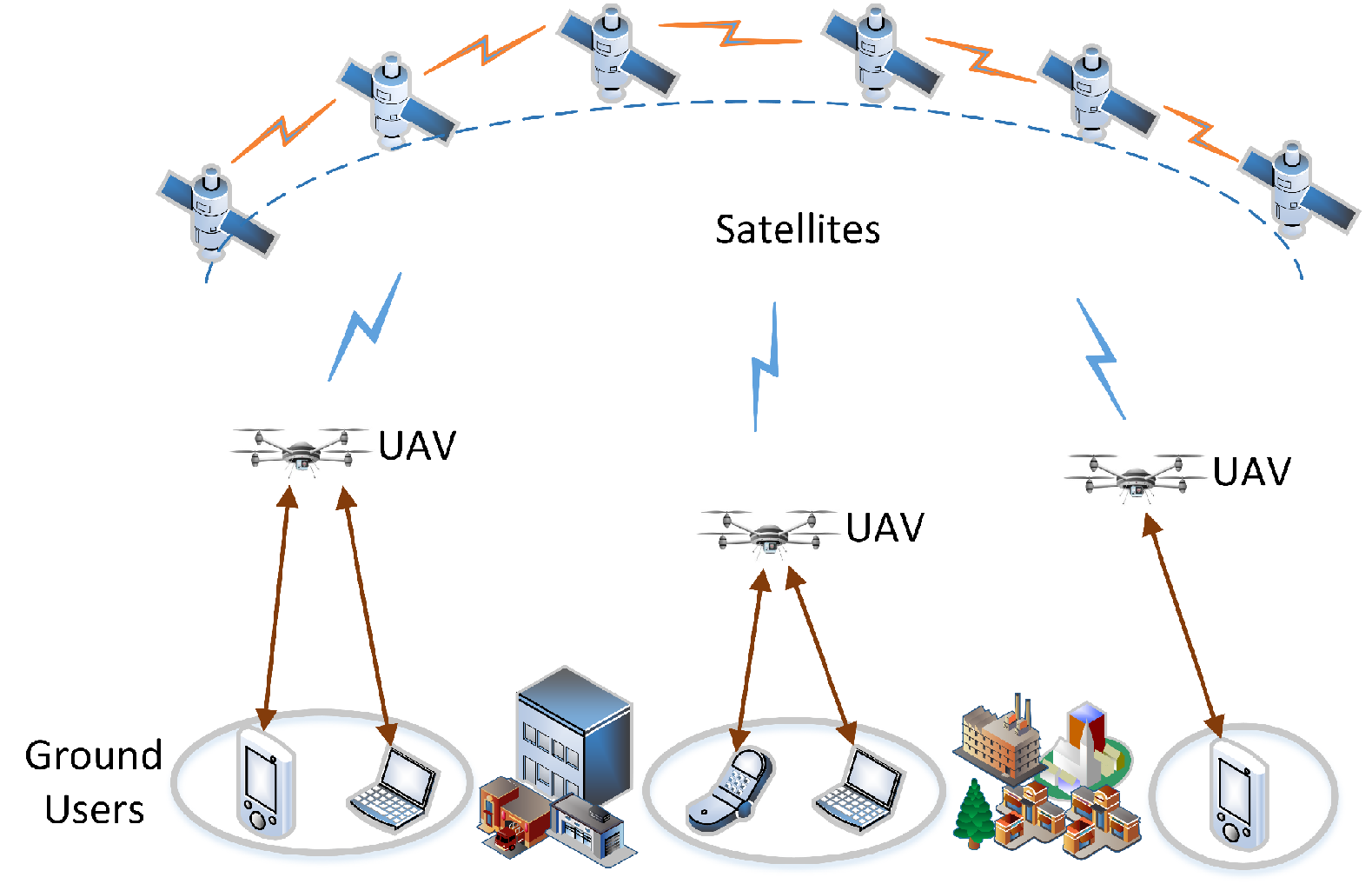}}
 \caption{System model of a SAGIN.} \label{fig:SM}
\end{figure}

\begin{table*}[t]
 \caption{Basic Symbol definitions for SAGINs}
  \centering
  \begin{tabular}{|c|c|c|c|}
  \hline
  \textbf{Definition} & \textbf{Symbol} & \textbf{Definition} & \textbf{Symbol}\\
  \hline
  Ground User&$G_k$&All UAV's Trajectory Set&$\mathcal Q$\\
  \hline
  UAV& $U_m$&UAV's Trajectory&$q_{m}$\\
  \hline
  LEO&$S_n$&Number of UAVs&$M$\\
  \hline
  Task&$\mathbb{T}_f$&Number of LEOs&$N$\\
  \hline
  User-UAV Cluster&$\wp$&  Number of Users&$K$\\
  \hline
  Distance&$d$& Number of Tasks&$F$\\
  \hline
  Dataset&$D$ &Number of Users&$K$\\
  \hline
  Bandwidth&$B$&User-UAV Pairing Indicator&$z^G$\\
  \hline
  Channel Rate&$C$&Number of UAVs Connected to Satellite $S_n$&$MS_{n}$\\
  \hline
  Rician Factor&$\omega$&Number of Users in Cluster $\wp_m$&$K_m$\\
  \hline
  AWGN&$\sigma$&UAV-Satellite Pairing Indicator&$z^U$\\
  \hline
  Antenna Gain&$\xi$&Path Loss Exponents(NLoS)&$\tau_N$ \\
  \hline
  Wavelength&$\lambda$&Path Loss Exponents(LoS)&$\tau_L$\\
  \hline
  Antenna Phase&$\varrho_m$&Edge Aggregation Weight&$\zeta^{f_K}$\\
  \hline
  Bessel Function&$\delta$&Cloud Aggregation Weight&$\zeta^{f_M}$\\
  \hline
  Thermal Noise&$\varphi$&Final Aggregation Weight&$\zeta^{f_S}$ \\
  \hline
  Boltzmann Constant&$\varsigma$&Transmit Power of LEOs&$P_{S}$\\
  \hline
  Height of LEOs&$R_S$ &Transmit Power of UAVs&$P_{U}$\\
  \hline
  Velocity of LEOs&$v_L$&Transmit Power of Ground Users&$P_{G}$\\
  \hline
  Final Aggregation LEO&$S^f$&User-UAV Pairing (Uplink)&$\Im_u$\\
  \hline
  Radius of the Earth&$R_E$&User-UAV Pairing (Dplink)&$\Im_d$\\
  \hline
  ISL Carrier Frequency&$F_{ISL}$&Minimum Elevation Angle&$\varpi$ \\
  \hline
  Normalized Gain of LEO&$\eta^d$&Peak Gain of LEO&$\eta_{\rm max}$\\
  \hline
 \end{tabular}
 \label{tab:pares}
\end{table*}

As shown in Fig. \ref{fig:SM}, we consider a three-layer SAGIN to provide global communication service. In the proposed SAGIN, there are $K$ ground users $G_k$ ($k \in \mathcal K = \{1, 2, ..., K\}$), $M$ UAVs $U_m$ ($m \in \mathcal M = \{1, 2, ..., M\}$) and $N$ low Earth orbit (LEO) satellites $S_n$ ($n \in \mathcal N = \{1, 2, ..., N\}$). UAVs can provide ground-to-air connectivity services for ground users and can also connect with satellites, each UAV can dynamically plan its trajectory to provide communication and computation services to ground users within its coverage area, we assume the 3D coordinates of UAV $U_m$ is $q_{m}(t) = \{x_{m}(t), y_{m}(t), z_{m}(t)\}$ ($q_m \in \mathcal Q = \{q_1, q_2, ..., q_M\}$)) at a given time slot $t$. However, due to the potential long distances between ground users, direct communication between them may not be feasible. Similarly, UAVs which provide services for ground users might not be able to communicate directly with each other. Furthermore, we consider that satellites are constantly in high-speed motion, thus the communication window between UAVs and LEO satellites is limited in time.

\subsection{Communication Model}
In this work, we assume that ground users are divided into $M$ clusters, a cluster $\wp_m$ ($m \in \{1, 2, ..., M\}$) has $K_m$ users, $\wp = \{\wp_1, \wp_2, ..., \wp_M\}$ denotes the cluster set. The cluster $\wp_m$ is serviced by the UAV $U_m$, and $z^G_{m,k}$ denotes ground user $G_k$ is assigned to the cluster $\wp_m$. Due to the considerable distance between different clusters, direct communication between them is not feasible. We assume the channels between UAVs and ground users follow Rician fading \cite{10312474}, and UAV $U_m$ serves cluster $\wp_m$ which includes users $G_k$, thus the channel coefficient $h_{m,k}$ between the UAV $U_m$ and the ground user $G_k$ is given as
\begin{equation}\small\label{eq:hmn}
h_{m,k}= \sqrt{\frac{\omega}{\omega+1}} \bar{H}_{m,k}+\sqrt{\frac{1}{\omega+1}} \hat{H}_{m,k},
\end{equation}
where $\omega$ indicates the Rician factor, $\hat{H}_{m,k}=\hat{\mathbb{g}}_{m,k} d_{m,k}^{-{\tau_N}/2}$ and $\bar{H}_{m,k}=\bar{\mathbb{g}}_{m,k} d_{m,k}^{-{\tau_L} / 2}$ indicate the non-line-of-sight (NLoS) and the line-of-sight (LoS) channel coefficients, respectively. $d_{m,k}$ denotes the distance between UAV $U_m$ and ground user $G_k$. $\tau_N$ and $\tau_L$ are the path loss exponents for NLoS and LOS channel, respectively. In the NLoS channel, the coefficient $\hat{\mathbb{g}}_{m,k}$ is formed by a zero-mean unit-variance Gaussian fading channel. Therefore, in a given cluster $\wp_m$, the transmission rate of the uplink channel from $G_k$ to $U_m$ can be expressed as
\begin{equation}\small\label{eq:Ratemn}
C_{m,k}= B_{k,m} {\rm{log_{2}}} \left(1 + \frac{P_{G_k} |h_{m,k}|^2} { \sum_{i=1,i \neq k}^{K} P_{G_i} |h_{m i}|^2 + {{\sigma}_{n}^2} } \right),
\end{equation}
where $B_{k,m}$ indicates the bandwidth of the channel from $G_k$ to $U_m$, $P_{G_k}$ denotes the transmit power of ground user $G_k$, ${{\sigma}_{n}^2}$ denotes the variance of the additive-white-Gaussian-noise (AWGN) at $U_m$. Similarly, the transmission rate of the downlink channel from $U_m$ to $G_k$ can be expressed as
\begin{equation}\small\label{eq:Ratemn2}
C_{k,m}= B_{m,k} {\rm{log_{2}}} \left(1 + \frac{P_{U_m} |h_{m,k}|^2} {{{\sigma}_{n}^2} }   \right),
\end{equation}
where $B_{m,k}$ indicates the bandwidth of the channel from $U_m$ to $G_k$, and $P_{U_m}$ denotes the transmit power of UAV $U_m$. Notice that in federated learning, the uplink and downlink transmissions occur asynchronously, and the edge server (e.g. $U_m$) broadcasts the aggregated model to all ground users in downlink transmissions. As such, there is no interference in this stage. On the other hand, we assume that a UAV can establish communication with only one LEO satellite during any given time slot, $z^U_{n,m}$ denotes UAV $U_m$ can communicate with satellite $S_n$ and $MS_{n}$ denotes the number of UAVs connected to satellite $S_n$. Thus, the channel coefficient between satellite $S_n$ and UAV $U_m$ can be expressed as \cite{9726800}
\begin{equation}\small\label{eq:G_s}
\hat{h}_{n,m}=\frac{\sqrt{\xi_m}\lambda}{4 \pi d_{n,m}} e^{j {\varrho_m}},
\end{equation}
where $\xi_m$ indicates the antenna gain, $\lambda$ is the wavelength, $d_{n,m}$ denotes the distance between the satellite and the UAV, $\varrho_m$ indicates the phase. In this work, we consider the impact of outdated channel state information (CSI) due to the considerable distance between satellites and aerial platforms. The outdated CSI is formulated as \cite{10287142}
\begin{equation}\small\label{eq:outdatedCSI}
h_{n,m}= \delta \hat{h}_{n,m} + \sqrt {1-\delta^2} \hat{g}_{n,m},
\end{equation}
where $\delta = \hat{\kappa} (2 \pi {\hat{D}}_{n,m} T_{n,m})$, $\hat{\kappa}$ denotes the Bessel function of the first kind of order 0, ${\hat{D}}_{n,m}$ and $T_{n,m}$ indicate the maximum Doppler frequency and the transmissions delay between the satellite and the UAV, respectively. $\hat{g}_{n,m}$ represents a complex Gaussian random variable with an equivalent variance to that of $\hat{h}_{n,m}$. Therefore, the transmission rate of the uplink channel from UAV $U_m$ to satellite $S_n$ is given as
\begin{equation}\small\label{eq:Ratenl}
C_{m,n}= B_{m,n} {\rm{log_{2}}} \left(1 +  \frac{P_{U_m} |h_{n,m}|^2} {\sum_{i=1,i \neq m}^{M} P_{U_i} |h_{n,i}|^2 + {{\sigma}_{l}^2} }   \right),
\end{equation}
where $B_{m,n}$ is the bandwidth of the channel from $U_m$ to $S_n$, $\sum_{i=1,i \neq m}^{M} P_{U_i} |h_{n,i}|^2$ is the interference from other UAVs, ${\sigma}_{l}^2$ is AWGN at $S_l$. Similarly, the transmission rate of the downlink channel from $S_n$ to $U_m$ can be expressed as
\begin{equation}\small\label{eq:Ratenl2}
C_{n,m}= B_{n,m} {\rm{log_{2}}} \left(1 +  \frac{P_{S_n} |h_{n,m}|^2} { {{\sigma}_{l}^2} }   \right),
\end{equation}
where $B_{n,m}$ is the bandwidth of the channel from $S_n$ to $U_m$, $P_{S_n}$ denotes the transmit power of satellite $S_n$ in downlink transmissions. Similar to the downlink transmission between UAVs and ground users, there is no interference at this stage. Moreover, there are inter-satellite links (ISLs) between satellites in current satellite networks, we thus also consider inter-satellite communications via ISLs. The transmission data rate of the ISL channel \cite{ISL_TWC21} between two satellites $S_{a}$ and $S_{b}$ can be expressed as
\begin{equation}\small\label{eq:rate_ISL}
C_{a b} = B_{a,b} {\rm{log_{2}}} \left( 1 +  \frac{P_{S_a} |\eta_{\rm max}|^2}{\varsigma \varphi B_{a b} (\frac{{4 \pi d_{a b} F_{ISL}}}{c})^2 } \right),
\end{equation}
where $B_{a,b}$ indicates the bandwidth of channel from $S_{a}$ to $S_{b}$, $P_{S_a}$ is the transmit power for the transmitter satellite $S_a$. $\eta_{\rm max} = {\rm max}_{J(a,b)} {\eta^d_{a,b}}$ represents the peak gain of the $S_{a}$ antennas in the direction of their main lobe, $J(a,b)$ denotes the relative direction between $S_{a}$ and $S_{b}$, $d_{a,b}$ indicates the distance $S_{a}$ and $S_{b}$, $\eta^d_{a,b}$ is the normalized gain of the satellite antennas in the transmission direction, $\varsigma$ denotes the Boltzmann constant, $\varphi$ is the thermal noise, $F_{ISL}$ denotes the ISL carrier frequency, and $c$ indicates the speed of light. It is noteworthy that interference can be omitted by ensuring that the inter-plane ISL antennas utilize sufficiently narrow beams along with precise beam steering or antenna pointing capabilities \cite{ISL_TWC21}.

\subsection{Satellite Coverage Model}
\begin{figure}[t!]
  \centering
  \centerline{\includegraphics[width=0.32\textwidth]{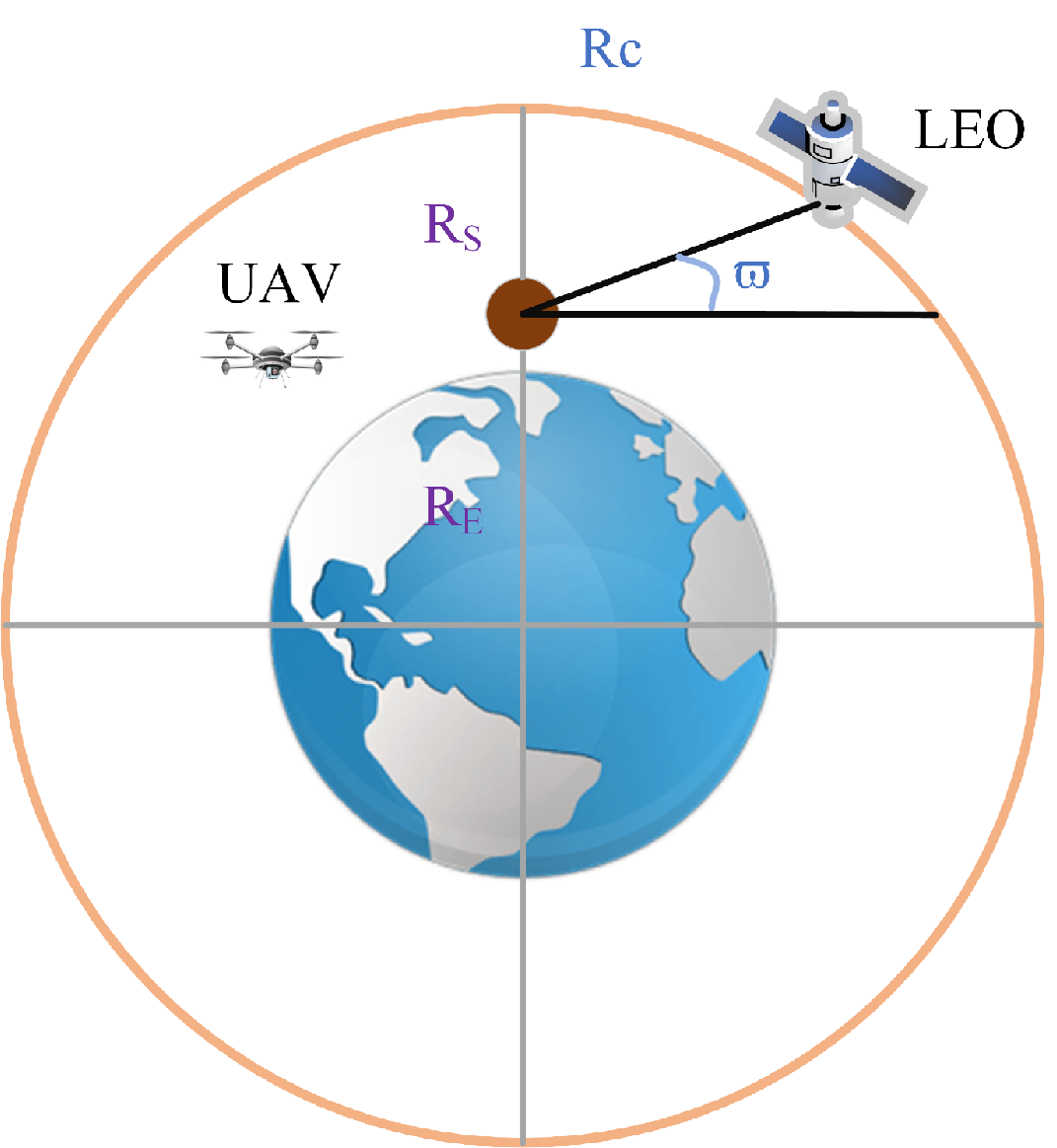}}
 \caption{The coverage area of a LEO satellite in SAGINs.} \label{fig:coverage}
\end{figure}

Considering that LEO satellites are constantly in high-speed motion, each LEO provides a limited time window for aerial (UAV) access. We consider the coverage area of satellites as illustrated in Fig. \ref{fig:coverage}, where the motion path length of an LEO satellite during the coverage access time for a specific UAV is as \cite{9344666}
\begin{equation}\small\label{eq:arcLength}
R_C = 2 (R_E + R_S) \bigg(\arccos  \Big(\frac{R_E}{R_E+R_S} \cos \varpi  \Big) - \varpi \bigg),
\end{equation}
where $R_E$ denotes the radius of the earth, $R_S$ indicates the height of the LEO satellite, $\varpi$ denotes the minimum coverage elevation angle for the LEO satellite. Therefore, we can obtain the total communication time for the LEO satellite and UAV as
\begin{equation}\small\label{eq:comTime}
T_c = \frac{R_C}{v_L},
\end{equation}
where $v_L$ denotes the orbital velocity of the LEO satellite. In this work, considering the spacing between LEO satellites, there are differences in the remaining service time between each LEO satellite and the same UAV at the beginning. Moreover, due to the negligible altitude of the UAV compared to the altitude of LEO satellites and the radius of the Earth, we assume the altitude of the UAVs can be disregarded in the computation of LEO coverage time window.

\subsection{Federated Learning Model}
\begin{figure}[t!]
  \centering
  \centerline{\includegraphics[width=0.5\textwidth]{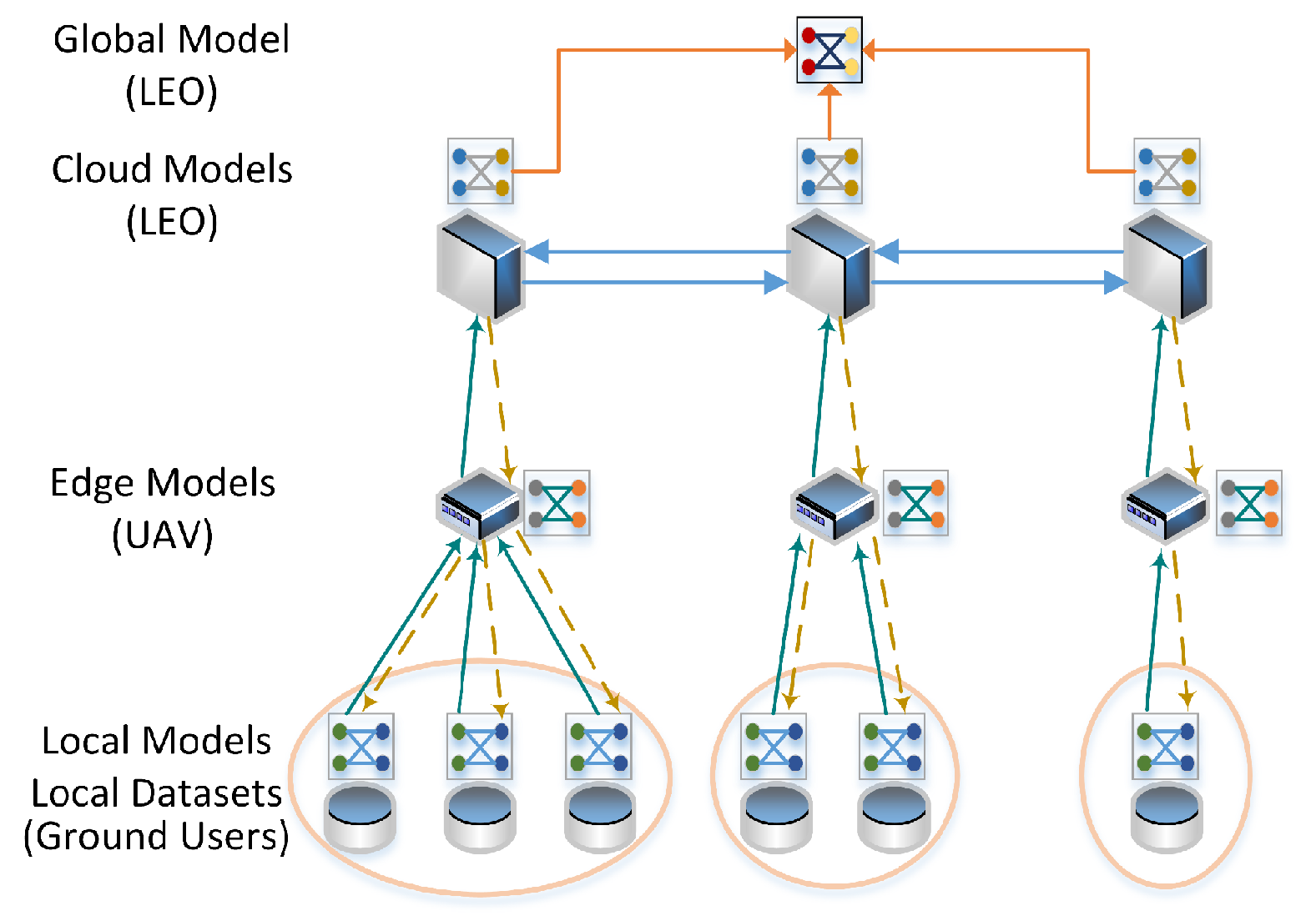}}
 \caption{The structure of the hierarchical federated edge learning in SAGINs.} \label{fig:fl}
\end{figure}
In the proposed federated learning framework, we assume there are $F$ distinct tasks $\mathbb{T}_f$ ($f \in {1, 2, ..., F}$) that need to be trained in the SAGIN. The local datasets of each ground user consists of $F$ datasets $D_k = \{D^1_k, D^2_k, ..., D^F_k\}$, corresponding to $F$ tasks. Notice that a zero-sized dataset indicates that the ground user cannot contribute to the corresponding task. For example, if $D^F_k$ is a zero-sized dataset, the ground user $G_k$ cannot participate in the training of task $\mathbb{T}_F$. Furthermore, we assume that each ground user can train only one iteration of one task per time slot. Therefore, the federated learning training process consists of four stages for a specific task $\mathbb{T}_f$: local training and edge aggregation, cloud aggregation and global update.

\subsubsection{Local Training and Edge Aggregation}
In the local training phase, user $G_k$ in cluster $\wp_m$ selects task $\mathbb{T}_f$ for training and updates its corresponding local model. Subsequently, user $G_k$ uploads the updated local model with the parameter $\mu^f_k$ to the corresponding UAV $U_m$. Considering the existence of multiple tasks, the local model being trained and the local model being uploaded may not belong to the same task. Therefore, the uploaded local model may have experienced multiple iterations of local training. We assume that the model size of task $\mathbb{T}_f$ is $DM_f$. However, due to the dynamic nature of wireless channels and the dynamic trajectory planning of UAVs, the uplink rates between ground users and UAVs are not constant in SAGINs which leads to variations in upload time as time varies. Therefore, UAV trajectory planning is crucial in this phase as it can mitigate the fluctuations in upload rate caused by channel variations. Subsequently, the edge server (UAV) $U_m$ will operate edge aggregation after receiving the local models from ground users within its served area. The aggregation equation is given by
\begin{equation}\small\label{FL_Aggregation}
\begin{aligned}
   \mu^f_m = \frac{1}{K_m}\sum_{i=1}^{K_m} \zeta^{f_K}_{i} \mu^f_i,
\end{aligned}
\end{equation}
where $K_m$ denotes the number of ground users in this cluster, $\zeta^{f_K}_{i}$ denotes the aggregation weight for the local model from ground user $G_i$. Due to the influence of the dynamic environment and different locations of ground users, the uplink transmission rate of ground users are different, thus the time required for uploading models are different among ground users. Furthermore, due to the presence of multiple tasks in federated learning, the number of training iterations experienced by the models uploaded by each ground user during each edge aggregation stage may vary. Moreover, considering the non-i.i.d. distribution of local datasets in a specific task, ground user's local datasets have different impact on the aggregated model. Therefore, the weight of local models in the aggregation is crucial to the result, and relying on traditional federated learning aggregation algorithms such as FedAvg's weighted average aggregation method can not fully optimize the model in SAGIN.

\subsubsection{Cloud Aggregation}
In this work, the LEO satellites are assumed to be the cloud servers for HFL in SAGINs. Each UAV can select one satellite within its service time for uploading. Each satellite receiving the edge model transmitted by UAVs will perform cloud aggregation, and then transmit it to a specific LEO satellite for final aggregation. The cloud model at LEO $S_i$ can be expressed as
\begin{equation}\small\label{FL_Aggregation2}
\begin{aligned}
   \mu^f_i = \frac{1}{M_i} \sum_{m=1}^{M_i} \zeta^{f_M}_{m} \mu^f_m,
\end{aligned}
\end{equation}
where $M_i$ denotes the number of UAVs which upload edge models to LEO $S_i$, $\zeta^{f_M}_{m}$ denotes the aggregation weight for the local model from UAV $U_m$. Thus, the final aggregation at LEO satellite $S_n$ can be expressed
\begin{equation}\small\label{FL_Aggregation3}
\begin{aligned}
   \mu^f_n = \frac{1}{M_n+N_n} \Big(\sum_{m=1}^{M_n} \zeta^{f_S}_{m} \mu^f_m + \sum_{j=1}^{N_n} \zeta^{f_S}_{j} \mu^f_j \Big),
\end{aligned}
\end{equation}
where $\mu^f_n$ denotes the global model, $M_n$ denotes the number of UAVs which upload edge models to LEO $S_n$, $N_n$ denotes the number of other LEO satellites which receive edge models from other UAVs, $\zeta^{f_S}_{j}$ denotes the aggregation weight for the local model from satellite $S_j$. Similar to the edge aggregation phase, the transmission rate between the air and space layer also varies due to the high-speed motion of satellites. Moreover, the model upload between UAVs and LEO satellites presents similar challenges as the previous phase, such as non-i.i.d. data distribution from ground users, different training iterations among different ground users, and differing aggregation iterations among different edge servers (UAVs).

\subsubsection{Global Update}
After an LEO satellite has aggregated all the models as the global model for a given task, it needs to broadcast the global model back to ground users. Considering the service time constraints of LEO satellites, an LEO with the global model may not be able to access all UAVs directly. In such cases, the global model may need to be transmitted to other satellites via ISLs and then relayed to corresponding UAVs. Subsequently, the UAVs transmit the global models to ground users within their coverage areas. Ground users continue to repeat the above steps after updating their local models by using the received global model until final convergence.

\subsection{Problem Formulation}
Our objective is to maximize the average performance (e.g. minimizing the loss in classification tasks) of all tasks while ensuring fairness among different tasks, i.e., reducing the discrepancy in performance across tasks. In the proposed SAGIN, each ground user is allocated in a cluster and may participate in one or more or even all tasks, while the service time of satellites is limited. Moreover, the dynamic nature of the wireless communication environment such as channel fading, affects the speed of model uploads and downloads. Therefore, it is necessary to plan the UAV trajectories to counteract the effects of channel fading and to optimize the clustering between ground users and UAVs. This strategy aims to enhance the communication efficiency between the ground layer and the aerial layer for the proposed HFL system.

Furthermore, UAVs can select whether all users within their cluster area participate in ground-aerial communications based on the progress of task training (fewer users can reduce interference). For example, if considerable progress is made in training a task, UAVs can prioritize communication with users associated with other tasks. Similarly, satellites can choose whether to access UAVs within their coverage area based on the progress of task training, considering that the data distribution of user groups served by different UAVs is uneven.

Moreover, the ISLs between satellites create a collaborative network for SAGINs. However, the selection of the LEO satellite for uploading models from UAVs and the selection of the final aggregation node between satellites also affects the total training rounds. Besides, within the limited satellite service time, multiple rounds of global aggregation need to be performed for each task, inevitably resulting in some satellites being unable to cover UAVs in the later stages. Therefore, selecting the final aggregation satellite node and utilizing ISL to reduce communication time and improve aggregation efficiency are crucial for HFL in SAGINs.

For the downlink transmission in the global update stage, the pairing issue between LEO satellites and UAVs also exists due to the variations in service time and rate caused by the high mobility of satellites. Furthermore, the trajectory of UAVs affects the efficiency of broadcasting the global model to ground users within their service clusters. Therefore, the pairing issue and the UAV trajectories are also crucial for the communication efficiency during the broadcasting stage.

Therefore, to fairly maximize the performance of all tasks within the limited service time of the LEO satellites, we need to optimize: 1) the trajectory of each UAV; 2) the pairing between UAVs and users (uplink); 3) the pairing between UAVs and satellites (uplink); 4) the selection of the final aggregation satellite, 5) the pairing between satellites and UAVs (downlink); 6) the weights for edge aggregation on UAVs; 7) the weights for cloud aggregation on satellites; and 8) the weights for final aggregation. We formulate the optimization problem as
\begin{align}
    \bold{\rm (P1)}: &\min_{\mathcal{Q}(t), \wp(t), \Im_u(t), S^f(t), \Im_d(t), \zeta^{f_{K}}(t), \zeta^{f_{M}}(t), \zeta^{f_{S}}(t)} \notag\\& \sum_{f=1}^{F}\sum_{t=1}^{T_f}\sum_{n=1}^{N}\sum_{m=1}^{M} {\mathcal L}_{k,f} (\psi_{k,f}, D_{k,f}),\label{SecrecyFunc}\\
    {\rm s.t.}&~ \sum_{m=1}^{M}K_m = K \tag{\ref{SecrecyFunc}{a}}, \label{SecrecyFuncSuba}\\
    &\sum_{m=1}^{M} z^G_{m, k} = 1, \forall k \in {\mathcal K} \tag{\ref{SecrecyFunc}{b}}, \label{SecrecyFuncSubb}\\
    &\sum_{n=1}^{N} z^U_{n, m} = 1, \forall n \in {\mathcal N} \tag{\ref{SecrecyFunc}{c}}, \label{SecrecyFuncSubc}\\
    &\sum_{n=1}^{N} MS_{n} = M \tag{\ref{SecrecyFunc}{d}}, \label{SecrecyFuncSubd}\\
    &\sum_{i=1}^{K_m} \zeta^{f_{K}}_i = 1, \forall \wp_m \in {\wp}  \tag{\ref{SecrecyFunc}{e}}, \label{SecrecyFuncSube}\\
    &\sum_{i=1}^{MS_{n}} \zeta^{f_{M}}_n = 1, \forall n \in {\mathcal N} \tag{\ref{SecrecyFunc}{f}}, \label{SecrecyFuncSubf}\\
    &\sum_{n=1}^{N} \zeta^{f_{S}}_n = 1  \tag{\ref{SecrecyFunc}{g}}, \label{SecrecyFuncSubg}\\
    &v_m \leq v_{\max}  \tag{\ref{SecrecyFunc}{h}}, \label{SecrecyFuncSubh}\\
    &z_m \geq z_{\min} ~\&~ z_m \leq z_{\max}  \tag{\ref{SecrecyFunc}{i}}, \label{SecrecyFuncSubi}
\end{align}
where $T_f$ denotes the number of training rounds in $f$-th federated learning task, $\Im_u(t) = \{\Im^u_1(t), \Im^u_2(t), ..., \Im^u_M(t)\}$ denotes the pairing between UAVs and satellites for uplink transmissions, $S^f$ denotes the final aggregation satellite, $\Im_d(t) = \{\Im^d_1(t), \Im^d_2(t), ..., \Im^d_M(t)\}$ denotes the pairing between UAVs and satellites for downlink transmissions. $\zeta^{f_{K}}$, $\zeta^{f_{M}}$ and $\zeta^{f_{S}}$ denotes the aggregation weights for edge aggregation, cloud aggregation and final aggregation, respectively. ${\mathcal L}_{k,f} (\psi_{k,f}, D_{k,f})$ denotes the local loss of $f$th task for ground user $G_k$ with local model parameter $\psi_{k,f}$ and local dataset $D_{k,f}$. \eqref{SecrecyFuncSuba} shows that all ground users are divided into $M$ clusters. \eqref{SecrecyFuncSubb} indicates that a ground user is assigned to only one cluster. \eqref{SecrecyFuncSubc} and \eqref{SecrecyFuncSubd} denote that each UAV need to access to one satellite. \eqref{SecrecyFuncSube}, \eqref{SecrecyFuncSubf} and \eqref{SecrecyFuncSubg} shows that the rule of the aggregation weights. \eqref{SecrecyFuncSubh} indicates the speed limitation of UAVs, and \eqref{SecrecyFuncSubi} presents the altitude constraint for UAVs.

The above optimization problem includes eight variables, comprising both continuous and discrete variables, and some optimization variables are interdependent (e.g., the UAV's trajectory and the UAV-ground user pairing). Thus, it is clear that this is a mixed-integer nonlinear optimization problem and is highly complex. Considering that our proposed system involves multi-round hierarchical aggregation in federated learning, it also represents a long-term planning problem. Traditional algorithms wouldn't be efficient in addressing this problem, and real-time computational complexity is a critical concern in dynamic wireless communication scenarios. To tackle this issue we employ a DRL algorithm which is well-suited for this kind of optimization problem and exhibits low computational complexity after deployment. Moreover, we explore a hybrid action space framework to enhance the DRL performance and consider the fairness among different tasks in reward design.

\section{Hybrid Action Space DRL Algorithm with Fairness Reward Function} \label{sec:DSAC}
The proposed problem in this work is a long-term optimization problem which can be effectively addressed by using DRL. However, the presence of several continuous and discrete optimization variables in the optimization problem makes DRL optimization difficult to converge. Some existing approaches tried to tackle this issue of DRL by discretizing or continuousizing all optimization variables. However, these methods sacrifice a portion of the optimization space, reducing the system's controllability, and may lead to challenges in exploring the huge action space when exploring different possibilities \cite{XU2023141}. Therefore, we employ a decoupling approach to separate the hybrid optimization variables and assign them to different agents within DRL for learning. Subsequently, to efficiently combine the optimization results from different agents, we utilize a maximum posteriori policy based algorithm to mix the discretized and continuous variables after optimization, providing an effective solution to the proposed optimization problem. Furthermore, considering the fairness among different tasks, we design a fair and effective reward scheme to ensure that each task achieves satisfactory training results.

\subsubsection{MDP Design in DRL}
We utilize a DRL algorithm named DSAC \cite{9448360} to address the optimization problem. Firstly, we need to model the proposed system as a Markov decision process (MDP). The MDP encompasses states, actions, and rewards. The state $s(t)$ at time slot $t$ in this paper is given by
\begin{equation}\small\label{eq:state}
\begin{aligned}
s(t) =&~\{\{h_{k, m}(t)\}_{k \in \mathcal K, m \in \mathcal M}, \{q_m(t)\}_{q_m \in \mathcal Q},\\
     &~\{T_{c(m,n)}(t)\}_{m \in \mathcal M, n \in \mathcal N}, {Acc}_i(t)_{i \in \{\mathcal M, \mathcal N\}}\},
\end{aligned}
\end{equation}
where $\{h_{k, m}(t)\}_{k \in \mathcal K, m \in \mathcal M}$ indicates the channel coefficients of ground-aerial links, $\{q_m(t)\}_{q_m \in \mathcal Q}$ denotes the UAV's trajectories, $\{T_{c(m,n)}(t)\}_{m \in \mathcal M, n \in \mathcal N}$ is the remain service time of space-aerial links, ${Acc}_i(t)_{i \in \{\mathcal M, \mathcal N\}}$ denotes the test performance in each aggregation node (UAVs and LEO satellites). Furthermore, the action $a(t)$ at time slot $t$ is expressed as
\begin{equation}\small\label{eq:action}
\begin{aligned}
a(t) = \{\mathbb{Q}(t), \wp(t), Im_u(t), S^f(t), Im_d(t), \zeta^{f_{K}}(t), \zeta^{f_{M}}(t), \zeta^{f_{S}}(t)  \},
\end{aligned}
\end{equation}
where $\mathbb{Q}(t)$ denotes the UAV trajectory planning, $\wp(t)$ is the pairing for ground users and UAVs in uplink transmissions, $Im_u(t)$ denotes the pairing for UAVs and LEO satellites in uplink transmissions, $S^f(t)$ denotes the selection of the final aggregation node, $Im_d(t)$ indicates the pairing for LEO satellites and UAVs in downlink transmissions. $\zeta^{f_{K}}(t)$, $\zeta^{f_{M}}(t)$ and $\zeta^{f_{S}}(t)$ presents the weights of models in edge, cloud and final aggregations, respectively.

Moreover, we need to design a reward function for the MDP to provide feedback to the agents of the DRL algorithm. Since our objective is to enhance the average performance of all tasks while ensuring the fairness, our reward design needs to consider fairness in task progress to guarantee that all tasks converge to satisfactory performance after the satellites service time ends. To this end, we design the reward function for $F$ tasks as :
\begin{equation}\small\label{eq:rewardFunction}
\begin{aligned}
r(t) = &~ \frac{\alpha}{F}\sum_{f=1}^{F} \frac{\gamma_f(t) / \epsilon_{c1}}{\epsilon_f } \\
&+ \frac{(1-\alpha)}{F}\sum_{f=1}^{F} \frac{\gamma_f(t) / \epsilon_{c2}}{\epsilon_f + | \frac{1}{F}\sum_{j=1}^{F}\gamma_j(t) / \gamma_f(t)  - 1|},
\end{aligned}
\end{equation}
where $\alpha$ denotes the time decay factor used to adjust the inevitable initial imbalance in task progress during access and the later requirement for overall balance, we define $\alpha = \beta^t$, where $\beta$ is the decay factor and $\alpha$ is set as 1 at the beginning. $\gamma_f(t)$ denotes the performance of task $\mathbb{T}_f$ at time slot $t$, $\epsilon_{c1}$ and $\epsilon_{c1}$ are constants used to normalize the rewards for tasks, $\epsilon_f$ is a constant introduced to prevent zero value of the denominator when the reward bias is zero.

The designed reward function can be divided into two parts. At the beginning of the federated learning training process, it's not feasible to achieve balance among all tasks when the proposed system starts operating. Therefore, the focus is more on optimizing the performance of all tasks. As time progresses, the emphasis of rewards gradually shifts towards the second part of the reward function. The designed reward bias ensures that tasks deviating too much from the performance mean receive certain penalties, while those approaching the performance mean contribute significantly to rewards. Thus, the designed reward function dynamically adjusts the reward values based on task performance and runtime, ensuring fairness among tasks while optimizing the performance of all tasks.

\subsubsection{DSAC-Based Optimization}
Existing research in reinforcement learning has started to shift towards distributional algorithm studies because distributional DRL can estimate the entire distribution of total returns rather than just the expected distribution. This is highly beneficial for the convergence of long-term rewards. However, the relationship between learning the return distribution and predicting values has not been discussed in traditional distributional DRL. Therefore, we introduce the DSAC algorithm which reduces the impact of overestimation by learning the distribution function of state–action returns. Firstly, the algorithm introduces the distributional idea into SAC, the return for state-action pairs in SAC can be represented as:
\begin{equation}\small\label{eq:dsac1}
\begin{aligned}
\theta^\pi(s(t),a(t)) = r(t) + \vartheta \iota(t+1),
\end{aligned}
\end{equation}
where $\pi$ denotes the policy in DRL, $\vartheta$ is the discount factor, $\iota(j) = \sum_{j=t}^{\infty} \vartheta^(j-t) [ r(j) - \nu {\rm log} \pi (a(j)|s(j)) ]$, where $\nu$ is an importance factor related to the entropy \cite{9448360}. The Q-value in DSAC can be expressed as
\begin{equation}\small\label{eq:dsac2}
\begin{aligned}
Q^{\pi}(s(t),a(t)) = \mathbb{E} \big[\theta^\pi(s(t),a(t))\big],
\end{aligned}
\end{equation}
where ${\mathbb E}[\cdot]$ indicates the expectation. Compared to the expected state-action returns in \eqref{eq:dsac2}, we can directly consider using soft returns in \eqref{eq:dsac1} to construct the algorithm. Considering the maximum entropy in this case, we can define the distribution of the Bellman operator as
\begin{equation}\small\label{eq:dsac3}
\begin{aligned}
\upsilon \theta^\pi(s(t),a(t)) = &~r(t) + \vartheta \big( \theta^\pi(s(t+1),a(t+1)) \\
&- \nu {\rm log} \pi (a(t+1)|s(t+1))\big).
\end{aligned}
\end{equation}
Then, we can update the soft return distribution based on \eqref{eq:dsac3} as
\begin{equation}\small\label{eq:dsac4}
\begin{aligned}
\hat{\theta}_{new} = {\rm argmin} \mathbb{E} \big[ d_D (\upsilon \hat{\theta}_{old}(\cdot|s(t),a(t)), \hat{\theta}(\cdot|s(t),a(t)) ) \big],
\end{aligned}
\end{equation}
where $d_D$ is the distance between new and old distributions which can be expressed by Kullback–Leibler (KL) divergence \cite{pmlrv70bellemare17a}. To update the soft return policy, we form the update function as
\begin{equation}\small\label{eq:dsac5}
\begin{aligned}
\Theta_{\hat{\theta}}(\chi) = - \mathop{\mathbb{E}}\limits_{\aleph_{\Theta_{\hat{\theta}}(\chi)}} \Big[{\rm log} \digamma \big(Psi^{\pi_{\phi^{\prime}}} \theta(s(t),a(t)) | \hat{\theta}(\cdot|s(t),a(t)) \big)   \Big],
\end{aligned}
\end{equation}
where $\chi$ and $\phi$ are the parameters of the distributional value function $\hat{\theta}(\cdot|s(t),a(t))$ and the agent’s behavior policy $\pi_\phi(\cdot|s)$, respectively. $\aleph_{\Theta_{\hat{\theta}}(\chi)} = (s(t),s(t+1),a(t),r(t))\sim \mathbb{B}, a(t+1)\sim \pi_{\phi_{\prime}}, \theta(s(t+1),a(t+1))\sim \hat{\theta}_{\chi^{\prime}}(\cdot|s(t+1),a(t+1))$, $\digamma$ is the probability distribution between states and actions, $\mathbb{B}$ denotes the experience buffer in DRL, $\chi_{\phi^{\prime}}$ and $\phi_{\phi^{\prime}}$ denote the parameters of the target return and target behavior policy, respectively. Then we can update $\chi$ by using the following equation
\begin{equation}\small\label{eq:dsac6}
\begin{aligned}
\nabla_\chi \Theta_{\hat{\theta}}(\chi) = &~- \mathop{\mathbb{E}}\limits_{\aleph_{\nabla_\chi \Theta_{\hat{\theta}}(\chi)}} \Big[\nabla_\chi {\rm log} \digamma \big(Psi^{\pi_{\phi^{\prime}}} \theta(s(t),a(t)) \\
&~| \hat{\theta}(\cdot|s(t),a(t)) \big)   \Big],
\end{aligned}
\end{equation}
where $\aleph_{\Theta_{\hat{\theta}}(\chi)} = (s(t),s(t+1),a(t),r(t))\sim \mathbb{B}, a(t+1)\sim \pi_{\phi_{\prime}}, \theta(s(t+1),a(t+1))\sim \hat{\theta}_{\chi^{\prime}}$. We assume $\hat{\theta}_{\chi}$ has a Gaussian distribution, thus we can obtain
\begin{equation}\small\label{eq:dsac7}
\begin{aligned}
\nabla_\chi \Theta_{\hat{\theta}}(\chi) = &~\mathop{\mathbb{E}}\limits_{\aleph_{\nabla_\chi \Theta_{\hat{\theta}}(\chi)}} \Big[ - \frac{\partial \Psi_{\hat{\theta}}(\chi)}{\partial Q_{\chi}(s(t),a(t))} \nabla_{\chi}Q_{\chi}(s(t),a(t)) \\
&~- \frac{\partial \Psi_{\hat{\theta}}(\chi)}{\partial \rho_{\chi}(s(t),a(t))} \nabla_{\chi}\rho_{\chi}(s(t),a(t)) \Big],
\end{aligned}
\end{equation}
where
\begin{equation}\small\label{eq:dsac8}
\begin{aligned}
\frac{\partial \Psi_{\hat{\theta}}(\chi)}{\partial Q_{\chi}(s(t),a(t))} = \frac{\upsilon^{\pi_{\phi_{\prime}}}\theta(s(t),a(t)) - Q_{\chi}(s(t),a(t)) }{\rho_{\chi}(s(t),a(t))^2},
\end{aligned}
\end{equation}
\begin{equation}\small\label{eq:dsac9}
\begin{aligned}
\frac{\partial \Psi_{\hat{\theta}}(\chi)}{\partial \rho_{\chi}(s(t),a(t))} = &~\frac{\upsilon^{\pi_{\phi_{\prime}}}\theta(s(t),a(t)) - Q_{\chi}(s(t),a(t)) }{\rho_{\chi}(s(t),a(t))^3} \\
&~- \frac{1}{\rho_{\chi}(s(t),a(t))},
\end{aligned}
\end{equation}
where $\rho_{\chi}(s(t),a(t))$ is a standard deviation of the Gaussian distribution for $\theta(\cdot|s(t),a(t))$. From \eqref{eq:dsac9}, it can be seen that the Q-value is very easy to be overestimated during updates. Moreover, when the standard deviation of the Gaussian distribution tends towards zero or infinity, gradient computation issues can arise. Therefore, we use the following equation to constrain the standard deviation
\begin{equation}\small\label{eq:dsac10}
\begin{aligned}
\rho_{\chi}(s(t),a(t)) = \max \Big( \rho_{\chi}(s(t),a(t)), \rho_{\min}  \Big),
\end{aligned}
\end{equation}
where $\rho_{\min}$ is the limitation factor. Another constraint method is to clip $\upsilon^{\pi_{\phi_{\prime}}}\theta(s(t),a(t))$ to control the range of the updated Q-values, with the clipping scheme as
\begin{equation}\small\label{eq:dsac11}
\begin{aligned}
\frac{\partial \Psi_{\hat{\theta}}(\chi)}{\partial \rho_{\chi}(s(t),a(t))} = &~\frac{\jmath - Q_{\chi}(s(t),a(t)) }{\rho_{\chi}(s(t),a(t))^3} \\
&~- \frac{1}{\rho_{\chi}(s(t),a(t))},
\end{aligned}
\end{equation}
where
\begin{equation}\small\label{eq:dsac12}
\begin{aligned}
\jmath = \ell \Big(\upsilon^{\pi_{\phi_{\prime}}}\theta(s(t),a(t)), Q_{\chi}(s(t),a(t)) - \emptyset, Q_{\chi}(s(t),a(t)) + \emptyset \Big),
\end{aligned}
\end{equation}
where $\ell[l,i,j]$ indicates $l$ is clipped between $[i, j]$, $\emptyset$ is the clipping factor. Subsequently, we adopt the policy update equation in DSAC as
\begin{equation}\small\label{eq:dsac13}
\begin{aligned}
\Theta_{\pi}(\phi) = \mathop{\mathbb{E}}\limits_{s(t)\sim \mathbb{B},a(t)\sim \pi_\phi} \Big[ Q_{\chi}(s(t),a(t)) - \nu {\rm log}(\pi_{\phi}(a(t)|s(t))) \Big].
\end{aligned}
\end{equation}
Then, we can update $\phi$ by
\begin{equation}\small\label{eq:dsac14}
\begin{aligned}
\nabla_\phi \Theta_{\pi}(\phi) = &~\mathop{\mathbb{E}}\limits_{s(t)\sim \mathbb{B},\Re} \Big[- \nu \nabla_\phi{\rm log}(\pi_{\phi}(a(t)|s(t))) \big(\nabla_{a(t)}Q_{\chi}(s(t),\\
&~ a(t))- \nu \nabla_{a(t)}{\rm log}(\pi_\phi(a(t)|s(t)))\big) \nabla_\phi \imath(\Re;s(t)) \Big],
\end{aligned}
\end{equation}
where $\imath(\Re;s(t)) = \hat{a} + \Re \bigodot \bar{a}$, $\Re$ is sampled from a fixed distribution, $\bigodot$ is Hadamard product, $\hat{a}$ and $\bar{a}$ are the mean and standard deviation of $\pi_{\phi}(\cdot|s(t))$, respectively.
\begin{figure}[t!]
  \centering
  \centerline{\includegraphics[width=0.34\textwidth]{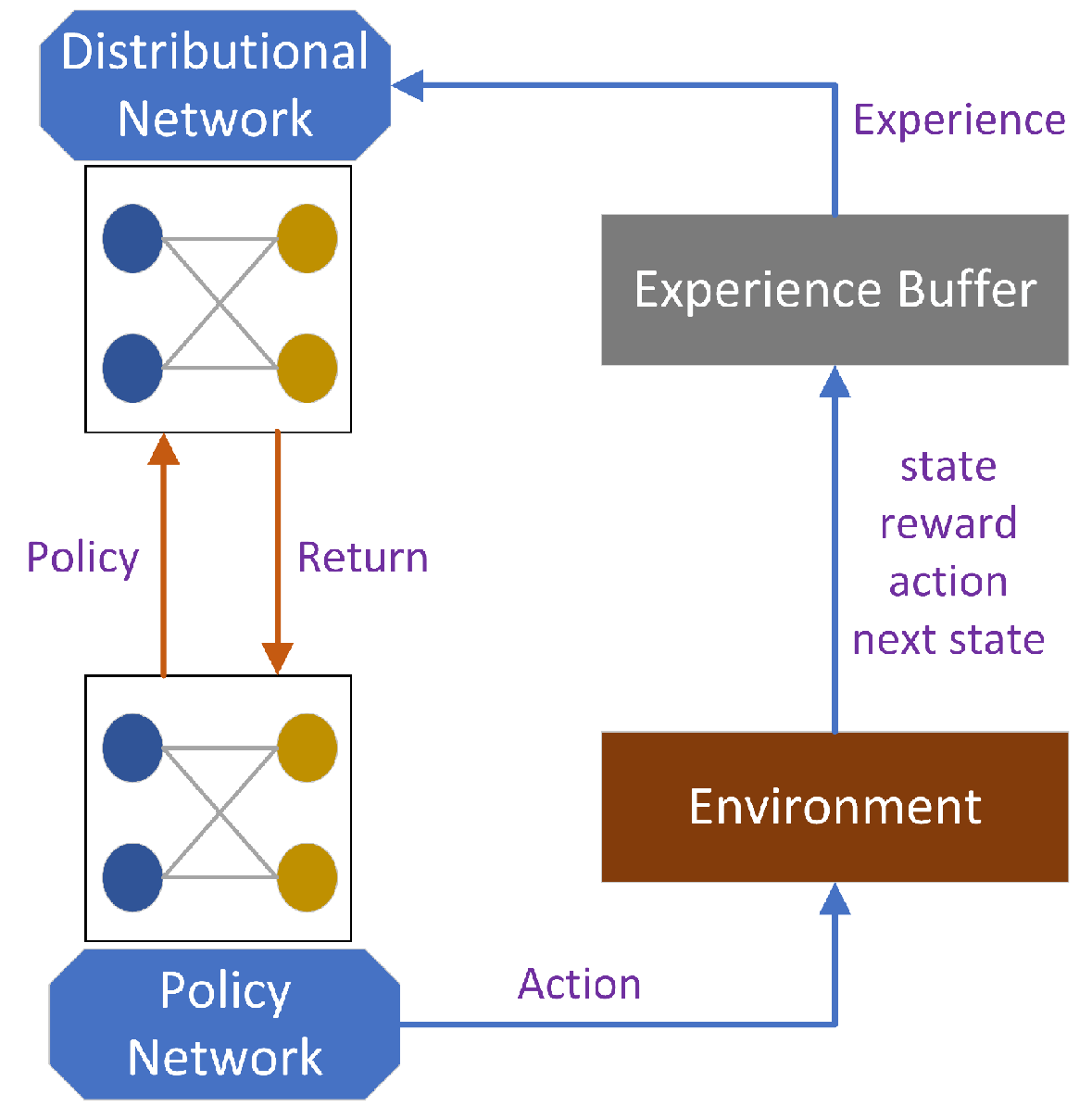}}
 \caption{The structure of DSAC.} \label{fig:DSAC}
\end{figure}
Moreover, we add a noise in the exploration of DRL as
\begin{equation}\small\label{eq:PolicyReparameteri}
\begin{aligned}
a(t) = V_{\phi}(\varepsilon(t); s(t)),
\end{aligned}
\end{equation}
where $V_{\phi}(i;j)$ indicates the random selection from noise and policy network, $\varepsilon$ is a Gaussian distributed noise. The structure of DSAC is shown in Fig. \ref{fig:DSAC}.

\subsubsection{Hybrid Action Space in DSAC}
\begin{figure}[t!]
  \centering
  \centerline{\includegraphics[width=0.46\textwidth]{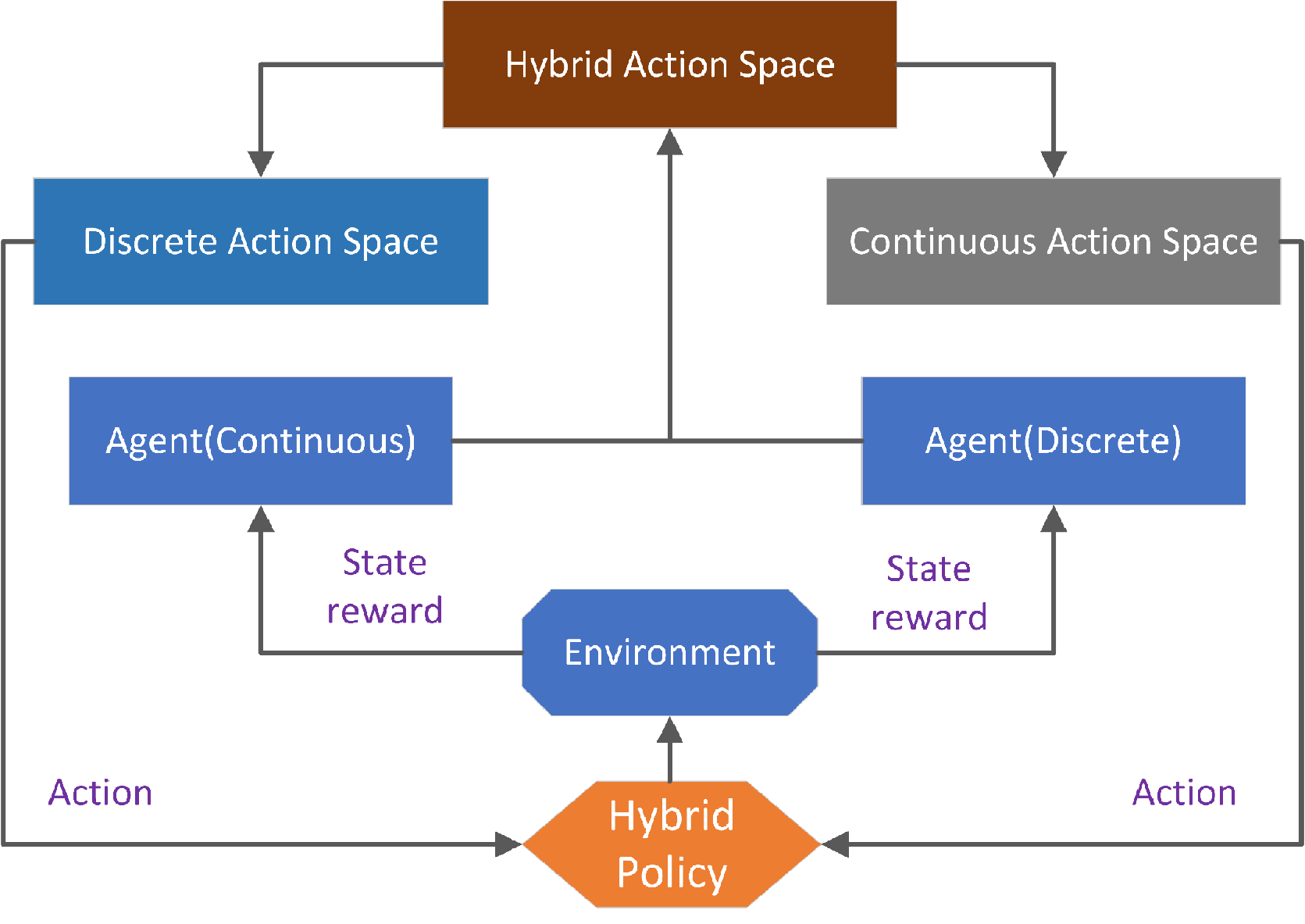}}
 \caption{The structure of hybrid action space in DSAC.} \label{fig:Hybrid}
\end{figure}

As mentioned in previous discussion, optimization problems that involve both discrete and continuous actions pose to a challenge to traditional DRL algorithms. Whether discretize or make all actions continuous, it leads to a decrease in system control capability and loss of optimization performance. Therefore, we introduce a decoupling strategy to decompose discrete and continuous actions in the optimization variables. Actions and policies are re-expressed as
\begin{align}
   A &~= \{A_d, A_c\}, \label{HAC_newAction}\\
   Q_\phi(a|s) &~= Q^{A_d}_\phi(a^{A_d}|s)Q^{A_c}_\phi(a^{A_c}|s) \notag\\
   &~= \prod_{a_i \in A_d}Q^{A_d}_\phi(a_i|s) \prod_{a_j \in A_c}Q^{A_c}_\phi(a_j|s), \label{AD_newPolicy}
\end{align}
respectively, where $A_d$ and $A_c$ indicate the discrete and continuous action space, respectively. Upon decomposing the optimization variables into discrete and continuous actions, we assign these two parts to different DRL agents for control, each based on an independently trained DSAC networks. Therefore, we formulate the update functions of value networks for discrete and continuous actions in DSAC networks as
\begin{equation}\small\label{HAC_cost1}
\begin{aligned}
    \Theta_{Q^{A_d}_{\chi}} = \mathop{\mathbb{E}}\limits_{(s(t),a(t)) \sim \mathbb{B}} \bigg[ \frac{1}{2} \Big({Q^{A_d}_{\chi}}\big(s(t), a(t)\big) - {Q}^{A_d}_{\chi^{\prime}}(s(t),a(t))\Big)^2  \bigg],
\end{aligned}
\end{equation}
\begin{equation}\small\label{HAC_cost2}
\begin{aligned}
    \Theta_{Q^{A_c}_{\chi}} = \mathop{\mathbb{E}}\limits_{(s(t),a(t)) \sim \mathbb{B}} \bigg[ \frac{1}{2} \Big({Q^{A_c}_{\chi}}\big(s(t), a(t)\big) - {Q}^{A_c}_{\chi^{\prime}}(s(t),a(t))\Big)^2  \bigg],
\end{aligned}
\end{equation}
respectively. The update function of policy networks for discrete and continuous actions can be expressed as
\begin{equation}\small\label{AD_cost3}
\begin{aligned}
    \Theta_{Q^{A_d}_{\phi}} =&~\mathop{\mathbb{E}}\limits_{s(t) \sim \mathbb{B}}\bigg[{\rm log}Q^{A_d}_{\phi}(V_{\phi}\big(\varepsilon(t); s(t))|s(t)\big)\\
    &- Q^{A_d}_{\chi}\big(s(t), f_{\phi}(\varepsilon(t); s(t))\big) \bigg],
\end{aligned}
\end{equation}
\begin{equation}\small\label{AD_cost4}
\begin{aligned}
    \Theta_{Q^{A_c}_{\phi}} =&~\mathop{\mathbb{E}}\limits_{s(t) \sim \mathbb{B}}\bigg[{\rm log}Q^{A_c}_{\phi}(V_{\phi}\big(\varepsilon(t); s(t))|s(t)\big)\\
    &- Q^{A_c}_{\chi}\big(s(t), f_{\phi}(\varepsilon(t); s(t))\big) \bigg],
\end{aligned}
\end{equation}
respectively. After decoupling the actions and obtaining samples from the environment separately by different agents, we need to re-couple the continuous and discrete action spaces. Therefore, we design a new policy $\phi_c$ for the proposed DSAC algorithm \cite{abdolmaleki2018maximum}, the update equation of $\phi_c$ can be expressed as
\begin{equation}\small\label{eq:ellUpdate}
\begin{aligned}
    \Theta_{Q_{\phi_c}} = \mathop{\mathbb{E}}\limits_{\phi_c(a(t)|s(t))}[\bar{Q}(s(t),a(t))],
\end{aligned}
\end{equation}
where $\bar{Q}$ indicates a new value network trained from the experience buffer \cite{neunert2020continuous}. When coupling the hybrid action space, we also need to consider the deviation during policy updates. Therefore, we can obtain
\begin{equation}\small\label{eq:ellUpdateKL}
\begin{aligned}
\mathop{\mathbb{E}}\limits_{s(t) \sim \mathbb{B}}[u(Q_{\phi_c}(a(t)|s(t))||Q_{\phi^{\prime}_c}(a(t)|s(t)))] < W,
\end{aligned}
\end{equation}
where $u$ indicates the KL divergence, $\phi^{\prime}_c$ denotes the old policy of $\phi_c$, and $W$ is a threshold factor to avoid deviation. The update function of the hybrid policy can be expressed as
\begin{align}
\hat{\phi}_c = ~&\arg \max_{\phi_c}\mathop{\mathbb{E}}\limits_{s \sim \mathbb{B}}[u(Q_{\phi_c}(a|s)||Q^{A_d}_{\phi_c}(a^{A_d}|s)Q^{A_c}_{\phi_c}(a^{A_c}|s))], \label{ellUpdate2}\\
{\rm s.t.}&~\mathop{\mathbb{E}}\limits_{s \sim \mathbb{B}} [u(Q^{A_d}_{\phi^{\prime}_c}(a^{A_d}|s)||Q^{A_d}_{\phi_c}(a^{A_d}|s))] < W_{A_d},\tag{\ref{ellUpdate2}{a}}, \label{ellUpdate2a}\\
&~\mathop{\mathbb{E}}\limits_{s \sim \mathbb{B}}[\frac{1}{Z}\sum_{z=1}^{Z} u(Q^{A_c}_{\phi^{\prime}_c}(a^{A_c}|s)||Q^{A_c}_{\phi_c}(a^{A_c}|s))] < W_{A_c} \tag{\ref{ellUpdate2}{b}}, \label{ellUpdate2b}
\end{align}
where $W_c$ and $W_d$ are the threshold for continuous and discrete action space, respectively. $Z$ represents the value of the discrete action space. Therefore, we employed a decoupling strategy to allocate the coupled optimization variables to different agents, and then coupled the optimized action spaces to form a hybrid solution. The proposed decoupling algorithm re-couples the variables after decoupling, providing DRL with stronger capabilities to explore the relationship between actions and states. Moreover, it is noticed that the prediction complexity in DRL is substantially lower than the training complexity. Once deployed, the pre-trained DRL model can make real-time decisions with very small computational overhead. The structure of the hybrid action space in DSAC method is shown in Fig. \ref{fig:Hybrid}. The pseudo-code of the proposed hybrid action space DSAC (H-DSAC) algorithm is in Algorithm \ref{Algorithm DSAC}.

\begin{algorithm}[t!]
\caption{\textbf{H-DSAC}: }\label{Algorithm DSAC}
\begin{algorithmic}[1]
 \makeatletter\setcounter{ALG@line}{0}\makeatother
 \State Initialize all the networks and parameters.
 \Repeat:
 \For {each time slot}
 \State Choose action $a(t)$ based on the current state $s(t)$
 \Statex \phantom{0}\phantom{0}\phantom{0}\phantom{0}\phantom{0}\phantom{0}and \eqref{eq:PolicyReparameteri}.
 \State Obtain the next state $s(t+1)$ and reward $r(t)$.
 \State Generate experience sample $\{s(t),a(t),r(t),$
 \Statex \phantom{0}\phantom{0}\phantom{0}\phantom{0}\phantom{0}\phantom{0}$s(t+1)\}$ and save it to the buffer $\mathbb{B}$.
 \EndFor
 \For {each training epoch}
 \State Sample a minibatch from the buffer $\mathbb{B}$.
 \State Update the importance factor $\nu \leftarrow \nu - L_{\nu}\nabla_{\nu}\Theta(\nu)$
 \State Update $Q^{A_d}_{\chi}$ based on \eqref{HAC_cost1}, $Q^{A_c}_{\chi}$ based on \eqref{HAC_cost2}.
 \State Update $Q^{A_d}_{\phi}$ based on \eqref{AD_cost3}, $Q^{A_c}_{\phi}$ based on \eqref{AD_cost4}.
 \State Soft update target networks $\chi^{\prime}$ and $\phi^{\prime}$.
 \State Update $\phi_c$ based on 1
 \EndFor
 \Until Convergence.
\end{algorithmic}
\end{algorithm}

\section{Simulation Results} \label{sec:sim}
In this section, we select 10 datasets (from Kaggle Computer Vision Open Datasets) as 10 tasks in federated learning and use TensorFlow-2 to train the tasks in simulations,
the size of the local dataset of a ground user for any given task is randomly generated. Unless otherwise stated, parameters in simulations are set as: the number of ground users $K = 10$, the number of UAVs $M = 3$, ground users and UAVs are randomly distributed within a 250 m × 250 m area. The ground users are randomly assigned to $M$ clusters, while the UAVs initially maintain an altitude of 50 meters. The maximum speed of UAVs $v_{\max} = 5$ m/s, $z_{\min} = 40$ m, $z_{\max} =$ 60m. The number of LEO satellites $N = 5$, the Rician factor $\omega = 10$ dB, the altitude of LEO satellites is 800 km, the speed of LEO satellites is 7.8 km/s, the minimum coverage elevation angle $\varpi = 40 ^{\circ}$, the carrier frequency for UAVs and LEOs are 1 GHz and 30 GHz, respectively. The antenna gain $\xi_m = 25$ dB for UAVs and 40 dB for LEOs, the maximum Doppler frequency is in the Ka-band \cite{10043628}. The bandwidth in ground-air-space transmissions is set as 10 MHz, the bandwidth for ISL transmissions is 1 GHz, the normalized gain $G^d = 1$ for the links between satellites, the thermal noise $\varphi = 354.81$ K \cite{ISL_TWC21}. The path loss exponents $\tau_L = 2$ and $\tau_N = 2.5$, the transmit power of ground users, UAVs and LEO satellites are 0.1 W, 1 W and 2 W, respectively. The discount factor $\vartheta = 0.99$, the decay factor $\beta = 0.995$, the thresholds $W = 0.1$, $W_c = 0.001$, $W_d = 0.01$, the normalizing constants $\epsilon_{c1} = 200$, $\epsilon_{c2} = 100$, $\epsilon_{f} = 0.01$, and $\rho_{\min} = 1$, the training time for each round is 1 s. We consider the distances between LEO satellites to be randomly generated between 100 km and 500 km. Moreover, we utilize parametrized DQN (PDQN) \cite{xiong2018parametrized}, multi-agent proximal policy optimization (M-PPO) \cite{MPPO}, `H-DSAC + FedAvg' (Using FedAvg \cite{FedAvg} to calculate the weights in aggregations) and `H-DSAC + HoveringUAV' (Without considering UAV trajectory planning, the UAVs remain in hovering states) to calculate the weights in aggregations as the benchmarks in simulations.

\begin{figure}[t!]
  \centering
  \centerline{\includegraphics[scale=0.55]{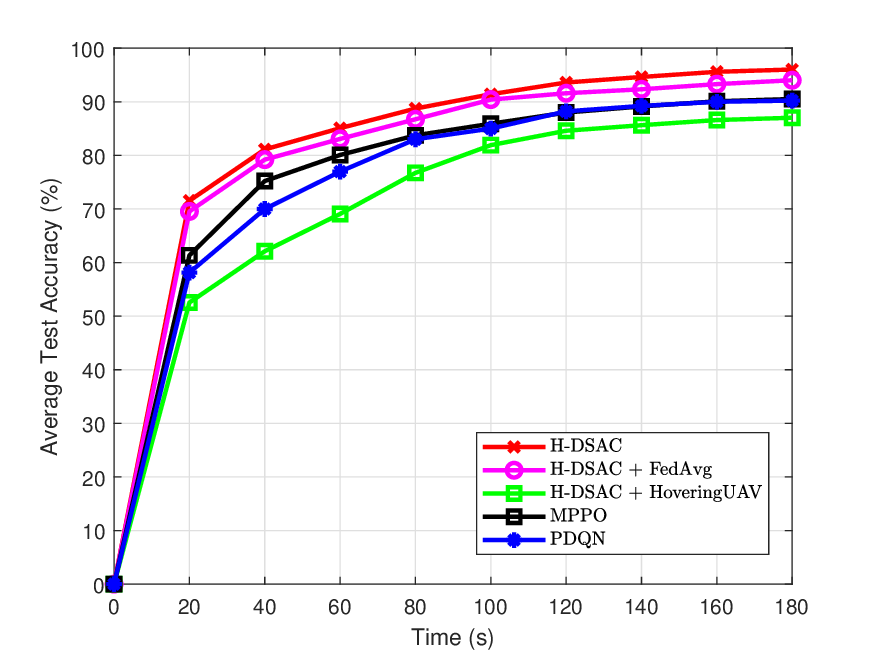}}
 \caption{\small Average test accuracy vs communication time.} \label{fig:R1}
\end{figure}

Fig. \ref{fig:R1} presents the average test accuracy of tasks versus the communication time for the proposed algorithm H-DSAC and benchmarks. As we can see in this figure, the average performance of tasks rises with an increase in the time. As the service time progresses, tasks continue to be trained in HFL which leads to the increase of accuracy. H-DSAC achieves the average test accuracy of 91.4\% when the communication time is 100 s, while the others achieve 90.4\%, 85.6\%, 84.9\% and 81.9\%, respectively. The proposed algorithm consistently outperforms all other algorithms, thus demonstrating the superiority of the H-DSAC algorithm. The performance of `H-DSAC + FedAvg' consistently ranks the second, indicating that the gain from the hybrid action space method is higher than the gain from aggregation weights in the proposed optimization problem. Compared to other algorithms, the performance of `H-DSAC + HoveringUAV' significantly deteriorates with the absence of UAV trajectory planning. Therefore, UAV trajectory planning plays a crucial role in ensuring the stability of ground-air links. In addition, MPPO and PDQN achieve similar performance across various scenarios, this result indicates that while MPPO may offer algorithmic superiority over DQN and DDPG, PDQN also possesses its own advantages in optimizing hybrid spaces.

\begin{figure}[t!]
  \centering
  \centerline{\includegraphics[scale=0.55]{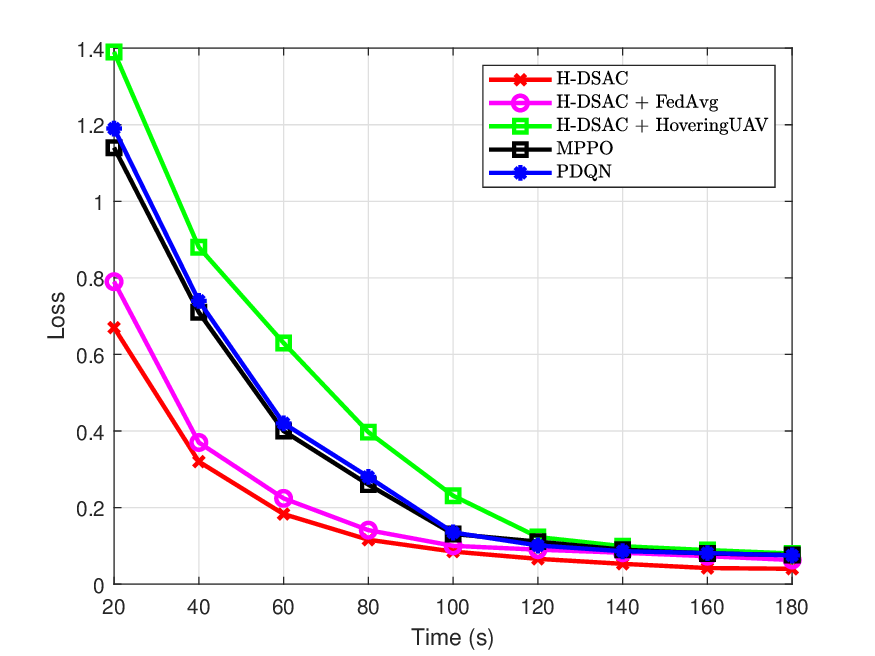}}
 \caption{\small Average loss vs communication time.} \label{fig:R1_loss}
\end{figure}

Compare the result in Fig. \ref{fig:R1_loss} to that in Fig. \ref{fig:R1}, it can be observed that while the loss is not directly related to the accuracy, their trends are very similar. As the average loss decreases continuously, the average accuracy increases correspondingly. H-DSAC achieves the average loss of 0.32 when the communication time is 40 s, while the others achieve 0.37, 0.88, 0.71 and 0.74, respectively. These results confirmed that the the proposed algorithm utilize the advantage of UAV trajectory control and weighted aggregation to enhance the performance. Moreover, The hybrid action space design offer algorithmic superiority over other DRL algorithms.

\begin{figure}[t!]
  \centering
  \centerline{\includegraphics[scale=0.55]{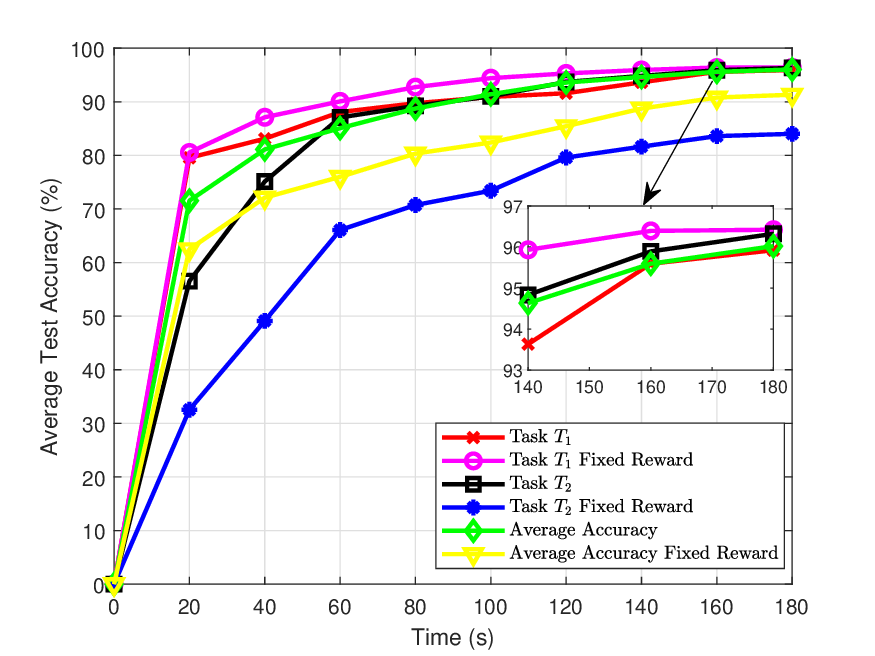}}
 \caption{\small Average test accuracy vs communication time.} \label{fig:R2}
\end{figure}

As illustrated in Fig. \ref{fig:R2}, we compare the performance of two selected task to show the fairness among tasks training, `Task $T_1$ Fixed Reward' refers to the performance of task $T_1$ during training, utilizing a fixed reward design instead of a dynamically adjusted reward function. It is clear that the performance of both two tasks rise rapidly at the beginning, Task $T_1$ achieves the average test accuracy of 89.7\% when the communication time is 80 s, while `Task $T_1$ Fixed Reward' achieves 90.7\%. It can be observed from the result that without a fairness-controlled reward function, certain tasks may converge to high performance more rapidly. Moreover, the fairness-oriented and dynamically adjusted reward function has small impact during the initial stages of communications based on \eqref{eq:rewardFunction}, resulting in slow progress for Task $T_2$ at the beginning. However, as the satellite communication time approaches its end, both tasks can converge to a high performance level with the assistance of the proposed dynamically adjusted reward function. Furthermore, we can see that without fairness considerations, Task $T_2$ may be consistently undervalued, leading to only around 82\% for testing accuracy even towards the end of communication. Besides, when we utilize the proposed reward function, the average accuracy is higher than that with fixed reward. This result underscores the beneficial role of the proposed reward mechanism for ensuring fairness among multiple tasks. If the satellite cannot guarantee that all tasks converge to a high accuracy threshold due to limited access time, the algorithm's dynamic design will ensure that all tasks fairly converge to the best possible value within the given time duration.

\begin{figure}[t!]
  \centering
  \centerline{\includegraphics[scale=0.55]{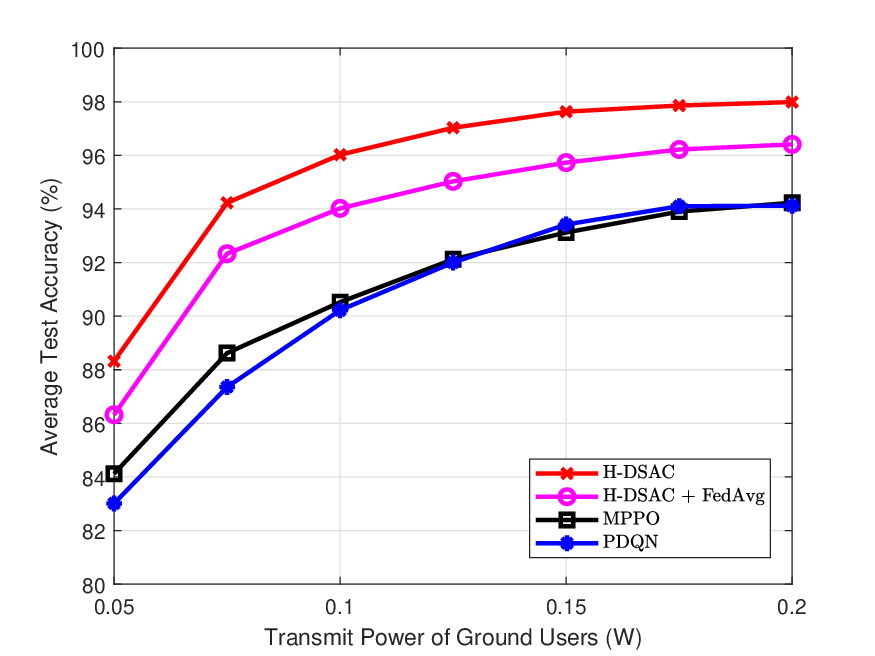}}
 \caption{\small Average test accuracy vs transmit power of ground users.} \label{fig:R3}
\end{figure}

Fig. \ref{fig:R3} compares the average test accuracy between the proposed method with different transmit power of ground users. It can be observed that the average test accuracy of tasks increases as the ground user power rises. This is because higher ground user transmission power can provide higher ground-to-air data rates, thereby reducing model transmission time and accelerating aggregation speed, allowing for more aggregation cycles within the limited satellite service time. Moreover, the proposed H-DSAC algorithm achieves around 97.99\% accuracy when the transmit power is 0.2 W for all ground users, while the other three benchmarks only achieve 96.4\%, 94.1\% and 94.2\%, respectively. This result demonstrates the superiority of the proposed algorithm over the benchmarks.

\begin{figure}[t!]
  \centering
  \centerline{\includegraphics[scale=0.55]{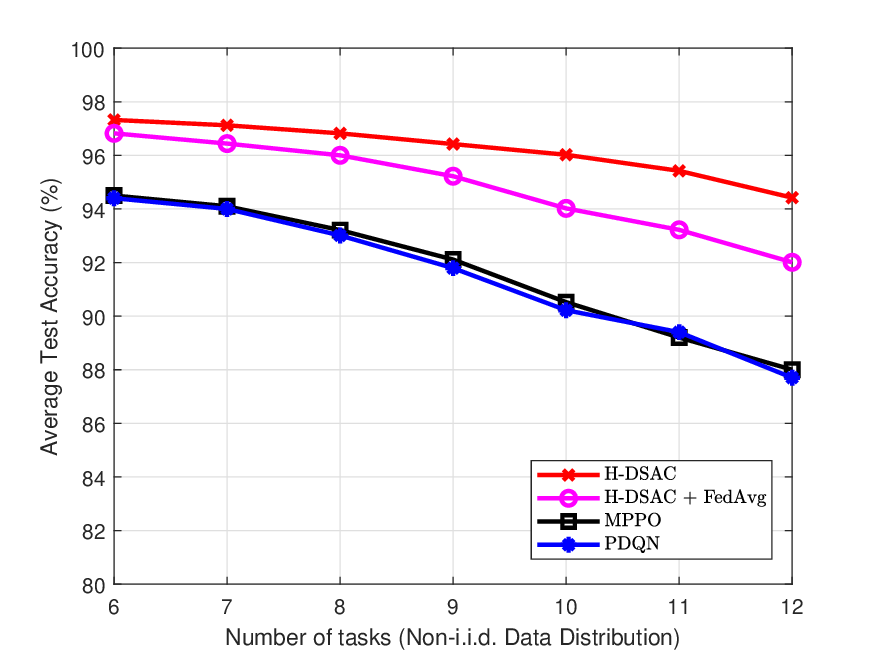}}
 \caption{\small Average test accuracy vs the number of tasks (non-i.i.d. data distribution).} \label{fig:R4}
\end{figure}

In Fig. \ref{fig:R4}, we compare the performance of federated learning tasks across different numbers of tasks, and consider that the data distribution for each task among different ground users follows non-i.i.d. From this figure, it can be seen that as the number of tasks increases, the average performance tends to deteriorate. This is because different tasks may originate from entirely different datasets, and an increase in the number of tasks inevitably leads to tighter constraints on transmission and computational resources. Moreover, the performance of the proposed algorithm consistently outperforms other benchmarks, H-DSAC achieves 96.9\% with 8 tasks, while others achieve between 91.7\% and 95.2\%. This result confirms the effectiveness of our proposed algorithm.

\begin{figure}[t!]
  \centering
  \centerline{\includegraphics[scale=0.55]{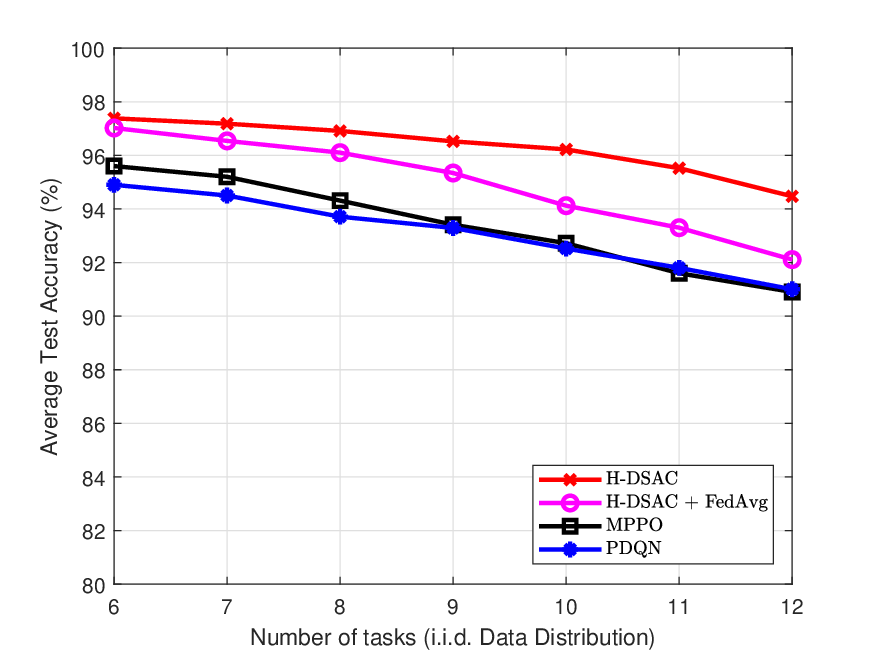}}
 \caption{\small Average test accuracy vs the number of tasks (i.i.d. data distribution).} \label{fig:R5}
\end{figure}

The results depicted in Fig. \ref{fig:R5} illustrate the performance of federated learning tasks when the dataset of each task follows i.i.d. among different ground users. Contrary Fig. \ref{fig:R4}, i.i.d. data facilitates to algorithm convergence because in this scenario, the importance of user models is equalized and each user holds an equivalent position in a specific task, necessitating only a balance in resource allocation to regulate their participation frequency in aggregation. Due to the inability of FedAvg to adjust the participation weights of users based on communication conditions for each aggregation, the proposed algorithm's advantage of adapting weights according to the state of communication links becomes more pronounced.

\begin{figure}[t!]
  \centering
  \centerline{\includegraphics[scale=0.55]{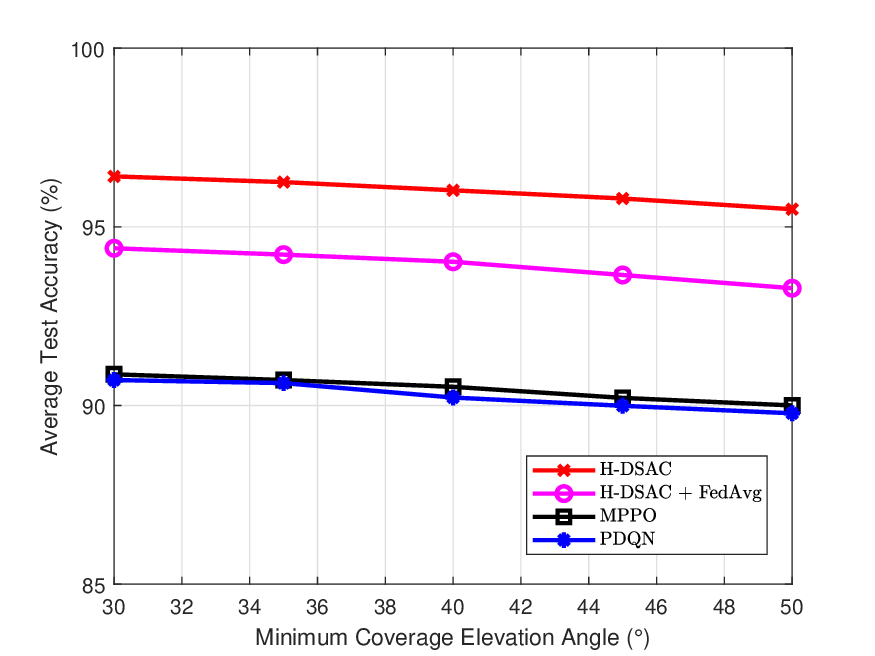}}
 \caption{\small Average test accuracy vs minimum coverage elevation angle.} \label{fig:R6}
\end{figure}

In Fig. \ref{fig:R6}, we assess the average test accuracy from both the proposed scheme and the benchmarks in relation to different minimum coverage elevation angle. As shown in all the results, the average test accuracy decreases as the minimum coverage elevation angle rises. The rationale is that, according to \eqref{eq:arcLength} and \eqref{eq:comTime}, as the minimum coverage elevation angle increases the service time between satellites and UAVs will decrease, which leads to a reduced service time for ground users to utilize SAGIN for task training. Therefore, both the number of federated learning training iterations and aggregation iterations will decrease, leading to a reduction in the average performance of all tasks.

\section{Conclusion}\label{sec:con}
In this paper, we proposed a novel application of the HFL framework within SAGIN, utilizing aerial platforms and LEO satellites as edge and cloud servers for HFL, respectively. We also considered ISLs as communication channels between satellites to further provide training resources for federated learning tasks. To maximize the average performance of multiple distinct tasks and ensure fairness among them during training, we employed a decoupling-coupling approach in the DSAC algorithm and designed a novel dynamically changing reward function to guarantee fairness among multiple tasks. By optimizing ground-to-air pairings in uplink and downlink transmissions, air-to-satellite pairings, aerial UAV trajectory planning, final aggregation selection between satellites, edge aggregation weights, cloud aggregation weights, and final aggregation weights, we balanced the training effects among different tasks to ensure that the performance of any individual task can converge to a high level while maximizing the overall performance. Through simulations, we validated the superiority of the proposed algorithm and emphasized the importance of fairness design during task training. Furthermore, we analyzed the impact of user transmit power, UAV trajectory planning and minimum coverage elevation angle on the results and demonstrated the significance of resource allocation in SAGINs for federated learning training performance. Moreover, we analyzed the effects of data distribution in tasks on algorithm convergence, the results show that the proposed algorithm exhibits strong adaptability to dynamic environments and provides a highly promising optimization tool for future federated learning frameworks and multi-access edge computing systems in SAGINs. In our future work, we will consider the adaptive downlink/uplink bandwidth allocation and full-duplex mode to further enhance the efficiency of federated learning in SAGINs.

\bibliographystyle{ieeetr}
\bibliography{ref}
\end{document}